\documentclass[fleqn,usenatbib]{mnras}

\usepackage{newtxtext,newtxmath}
 
\usepackage{mdframed}
\usepackage{hyperref}

\usepackage[T1]{fontenc}
\usepackage{ae,aecompl}

\usepackage{graphicx}	
\usepackage{amsmath}	
\usepackage{amssymb}	
\usepackage{todonotes}
\usepackage{ulem}
\usepackage{soul}
\usepackage{float}
\usepackage{multirow}
\usepackage[displaymath, mathlines,running]{lineno}

\usepackage[capitalize]{cleveref}
\Crefname{equation}{Equation}{Equations}
\crefname{equation}{Eq.}{Eqs.}
\Crefname{figure}{Figure}{Figures}
\crefname{figure}{Fig.}{Figs.}
\Crefname{table}{Table}{Tables}
\crefname{table}{Tab.}{Tabs.}
\Crefname{section}{Section}{Sections}
\crefname{section}{Sec.}{Secs.}

\def\LCDM{$\Lambda$CDM}
\def\Planck{\textit{Planck}}
\def\Z{\mathcal{Z}}
\def\P{\mathcal{P}}
\def\D{\mathcal{D}}
\def\L{\mathcal{L}}
\def\rmd{\mathrm{d}}
\def\var{\mathrm{var}}
\newcommand{\kmsMpc}{{\,\rm km/s/Mpc}}

\definecolor{ForestGreen}{RGB}{36,179,0}

\let\oldequation\equation
\let\oldendequation\endequation

\renewenvironment{equation}
  {\linenomathNonumbers\oldequation}
  {\oldendequation\endlinenomath}
  
\title[Assessing tension metrics with DES and Planck]{Assessing tension metrics with Dark Energy Survey and Planck data}

\author[DES Collaboration]{
\parbox{\textwidth}{
\Large
P.~Lemos,$^{1,2}$
M.~Raveri,$^{3}$
A.~Campos,$^{4}$
Y.~Park,$^{5}$
C.~Chang,$^{6,3}$
N.~Weaverdyck,$^{7}$
D.~Huterer,$^{7}$
A.~R.~Liddle,$^{8,9,10}$
J.~Blazek,$^{11,12}$
R.~Cawthon,$^{13}$
% ...
\begin{center} (DES Collaboration) \end{center}
}
}

\usepackage{eso-pic}

\AddToShipoutPictureBG*{%
  \AtPageUpperLeft{%
    \hspace{0.75\paperwidth}%
    \raisebox{-3.5\baselineskip}{%
      \makebox[0pt][l]{\textnormal{DES-2019-0456}}
}}}%

\AddToShipoutPictureBG*{%
  \AtPageUpperLeft{%
    \hspace{0.75\paperwidth}%
    \raisebox{-4.5\baselineskip}{%
      \makebox[0pt][l]{\textnormal{FERMILAB-PUB-20-662-AE}}
}}}%

\date{Accepted XXX. Received YYY; in original form ZZZ}

\pubyear{2020}

\begin{document}
\label{firstpage}
\pagerange{\pageref{firstpage}--\pageref{lastpage}}
\maketitle

\begin{abstract}
Quantifying tensions --- inconsistencies amongst measurements of cosmological parameters by different experiments --- has emerged as a crucial part of modern cosmological data analysis.
Statistically-significant tensions between two experiments or cosmological probes may indicate new physics extending beyond the standard cosmological model and need to be promptly identified.
We apply several tension estimators proposed in the literature to the Dark Energy Survey (DES) large-scale structure measurement and \Planck\ cosmic microwave background data. 
We first evaluate the responsiveness of these metrics to an input tension 
artificially introduced between the two, using synthetic DES data.
We then apply the metrics to the comparison of \Planck\ and actual DES Year 1 data. 
We find that the parameter differences, Eigentension, and Suspiciousness metrics all yield similar results on both simulated and real data, while the Bayes ratio is inconsistent with the rest 
due to its dependence on the prior volume. Using these metrics, 
we calculate 
the tension between DES Year 1 $3\times 2$pt and \Planck, finding the surveys to be in $\sim 2.3\sigma$ tension under the $\Lambda$CDM paradigm. This suite of metrics provides a toolset 
for robustly testing tensions in the DES Year 3 data and beyond.
\end{abstract}




%
\section{Introduction}
Two experiments are generally expected to agree, roughly within the reported errors, on the measured values of cosmological parameters. A disagreement between such measurements --- a \textit{tension} --- may be a sign of a mistake in one or both analyses, of unaccounted-for systematic errors, or perhaps of new physics. 
A prominent historical example of such tensions in cosmology is the disagreement between a variety of measurements of the matter density $\Omega_{\rm m}$ in the 1980s and 1990s that was vigorously debated at the time~\citep{Peebles:1984, Efstathiou:1990, Ostriker:1995, Krauss:1995yb} and eventually turned out to be explained by the discovery of the accelerating universe~\citep{Perlmutter:1998np, Riess:1998cb}. 

Presently, the discrepancy between the measurements of the Hubble constant using the distance ladder, $H_0 = (74.03 \pm 1.42) \kmsMpc$ \citep{Riess:2019cxk}, and those from \Planck, $H_0 = (67.4 \pm 0.5) \kmsMpc$ \citep{PlanckParameters:2018}, is much discussed, as it may be a harbinger of new physics.
Similarly, recent measurements of the parameter combination\footnote{
Here $\sigma_8$ is the present-day linear theory root-mean-square amplitude of the matter fluctuations averaged in spheres of radius $8 \ h^{-1} \ {\rm Mpc}$.} $S_8 \equiv \sigma_8(\Omega_{\rm m} / 0.3)^{0.5}$ from large-scale structure by the Dark Energy Survey \citep[DES, ][]{DES-3x2:2018} and the Kilo Degree Survey \citep[KiDS, ][]{Asgari:2020, Heymans:2020} differ from the cosmic microwave background (CMB) estimates
from the \Planck\ satellite at $\sim 2$--$3\sigma$ significance. 
These $N\sigma$ quantifications of tension are generally understood to correspond to probabilities equivalent to one-dimensional normal distribution, so that $1\sigma$ corresponds to 68\% confidence that the measurements are discrepant, $2\sigma$ corresponds to 95\%, etc.

The challenge is how to convert constraints from two data sets into such a probabilistic measure of tension between them. There exist a variety of methods to do this, which are being actively used in the community.   
While these \textit{tension metrics} are expected to give consistent messages in cases where the two data sets obviously agree or disagree, in more marginal cases the differences amongst them
--- including how much they depend on an analysis' choice of priors, assumptions of posterior Gaussianity, and the higher-dimensional shape of the posterior ---- have the potential to alter the assessment of whether or not two data sets are in agreement.

In the lead-up to cosmological results expected from the analysis of DES year 1 to year 3 data (henceforth simply Y3) 
and to inform other future cosmological analyses, we wish to provide a comprehensive characterization of how several proposed methods compare to one another. We also wish to confront these results with our intuition for what these metrics ought to be telling us about the agreement or disagreement between measurements. We specifically apply the methods to assess the consistency of DES and \Planck. This paper complements two earlier analyses that test the consistency of probes within DES \citep{Miranda:2020lpk,Doux:2020kdz}. 

These metrics serve only as diagnostics for whether there is tension, and not as a solution. If tension exists, it would indicate either unaccounted-for systematic effects in one or both experiments, or that the underlying model is inadequate to explain the data. 

Our basic approach is to create a suite of simulated DES data sets with a controlled level of induced tension relative to the best-fit \Planck\ 2018 cosmology. We then apply a number of methods to quantify this synthetic tension and assess their performance. Finally we apply the same tension metrics to quantify any tension between the published constraints from the first year of DES data (DES Y1)
and the \Planck\ 2015 and 2018 data sets.

The paper is structured as follows: 
We discuss the difficulties of tension estimation, and present the motivation of the present problem in \cref{sec:motivation}. We then describe our methodology in \cref{sec:simulated}. 
The different tension metrics studied in this paper are presented in \cref{sec:metrics}. 
We show results on simulated DES data in \cref{sec:results}, apply the tension metrics to DES Y1 in \cref{sec:y1}, and present our conclusions in \cref{sec:conclusions}.

\section{Motivation} 
\label{sec:motivation}
\begin{figure}
	\includegraphics[width=\columnwidth]{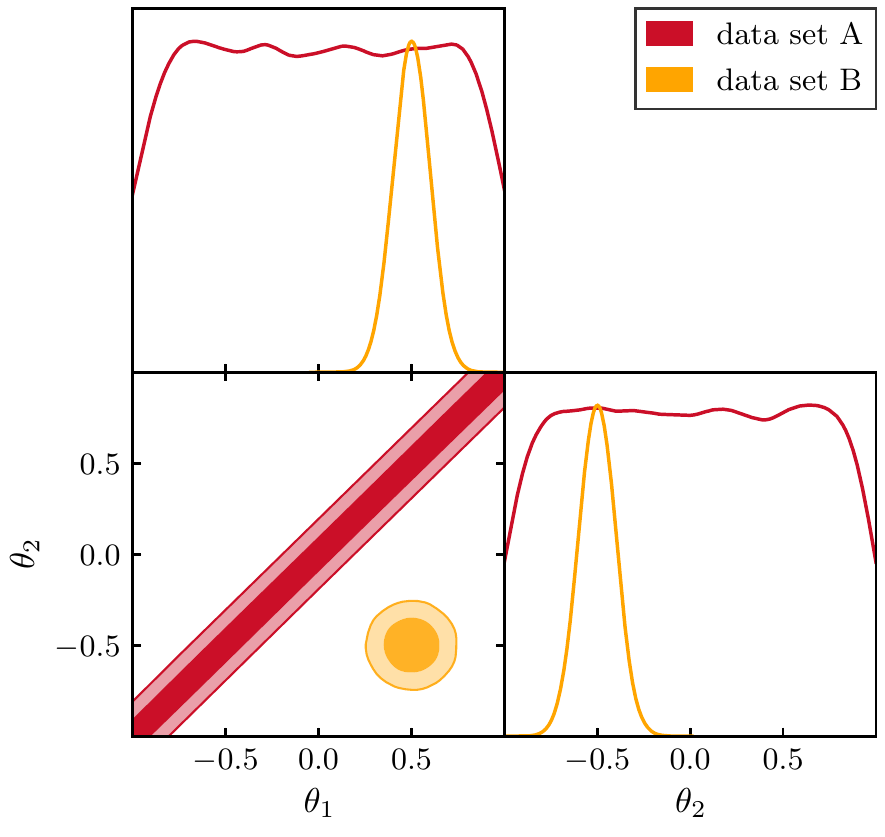}
    \caption{ \label{fig:example}
    Toy model example of a set of 2D constraints, where the 1D projections hide the discrepancy between the two data sets.
    The darker and lighter shade correspond to the $68\%$ and $95\%$ confidence regions respectively.}
\end{figure}

For a tension in a single parameter with an approximately Gaussian posterior distribution, it is easy to define a robust tension metric, as one can just report the one-dimensional difference between the posterior means 
of the two measurements divided by the quadrature sum of the errors reported by the two experiments.
For example, if \Planck\ reports that 
$S_8 =0.832\pm0.013$~\citep{PlanckParameters:2018} 
and DES reports $S_8=0.782\pm 0.022$~\citep{Troxel:2017xyo}, then one simply adds the errors in quadrature and reports the two results to be different at the level of
\begin{equation}
\label{eq:nsigma_1d}
\frac{\Delta S_8}{\sigma_{S_8}} = \frac{0.832-0.782}{\sqrt{0.013^2+0.022^2}}=2.0
\end{equation}
standard deviations, that is, they are in tension at the 
$2 \sigma$ level. 
However, as soon as we consider a tension in two or more parameters, this simple procedure becomes inadequate because full two-dimensional information cannot be captured by its one-dimensional projections. 
\cref{fig:example} gives an example showing how this intuition breaks down when the parameter space is multi-dimensional. If one were to judge consistency between the two data sets solely through their marginalized 1D constraints, one would conclude that the two data sets are consistent with each other. However, as evident from the comparison of their full 2D parameter constraints, the two data sets are in strong tension. Further complications arise when, for instance, one or more of the posteriors are non-Gaussian, or when the two posteriors originate from different prior assumptions on the parameters of interest.

There is no unique, universally-accepted method to quantify tension under these complicating circumstances. A variety of methods have been proposed, reviewed and tested
\citep{Charnock:2017vcd}.
Given this array of options, it is not obvious what the best choice is for a given analysis. In order to aid in this determination, in this paper we will describe and study several of these methods in order to compare their performance when applied to DES data. In doing so, we distinguish between two kinds of tension:
\begin{enumerate}
    \item \textbf{Internal} tensions, between different cosmological probes within one experiment (e.g.\ DES cosmic shear vs.\ galaxy clustering within DES).
    \item \textbf{External} tensions, between different experiments (e.g.\ DES vs.\ \Planck). 
\end{enumerate}
These must be treated differently because data-related systematic effects within the same experiment are often strongly correlated, necessitating use of more complex statistical tools when studying consistency.
While our methodology can be applied to either type of tension, here 
we specifically apply it to the case of external tensions. In addition, we focus on quantifying the tension between the large-scale structure measurements (via the combination of galaxy clustering, galaxy--galaxy lensing and cosmic shear, or often referred to as the ``$3\times 2$pt'' probes) from DES, and the CMB measurements from \Planck. 
Internal tension will be separately and additionally studied in~\citet{Doux:2020kdz} using Posterior Predictive Distributions (PPD) \citep{Gelman:2013}, which allow us to quantify tension in the presence of correlated systematic errors in the data, and to visualize the source of tension in the data vector. 
We do not consider the PPD in this work since it is not well suited to external tensions where there are many parameters that the two data sets do not share.

The challenge of accurately quantifying tension starts to become apparent as we investigate the expected performance of the tension metrics. Na\"{\i}vely, one might think that shifting one parameter by a controlled number of marginalized $N$ standard deviations would imply that the tension in the full-dimensional space would also be $N \sigma$; or in other words, that the amount of tension in the full, $N$-dimensional space is equal to the tension projected\footnote{In this paper the terms `marginalized over' and `projected' both mean `integrated over the other parameters'.}
to the original dimension.
However, this is not the case, because of two effects:
\begin{itemize}
    \item Marginalization 
    can hide tension that can only be seen in higher dimensions. This is caused by the fact that marginalisation leads to loss of information. This means that the full-dimensional tension can be larger than that inferred by looking at 1D distributions of the parameters.  
    This is illustrated with the simple two-dimensional example shown in \cref{fig:example}: 
    there are two parameters $\theta_1$ and $\theta_2$, and they are highly correlated as measured by experiment 1, but largely uncorrelated as measured by experiment 2. Because experiment 1 determines both parameters separately quite poorly, one-dimensional plots of the posterior show general agreement between measurements of the two experiments. Yet the two-dimensional plot shows that the two contours are significantly separated. This is because the well-measured combination of $\theta_1$ and $\theta_2$ significantly differs between experiment 1 and experiment 2.
    \item Relatedly, the number of dimensions of the problem also affects the inferred tension. The significance of a difference in parameter estimations between two experiments depends on the number of parameters constrained simultaneously by both experiments. Consider, for example, two experiments that measure the same parameter $\theta$ and obtain a one-dimensional $3 \sigma$ disagreement. The level of significance of this result is much higher if $\theta$ is the only parameter constrained by both experiments, than it is if the experiments also measure a hundred extra parameters, with no significant discrepancies between them. This common problem of the dilution of true tension with multiple comparisons is well known in statistics. For example, \cite{Heymans:2020} report a $\sim 3\sigma$ tension with \Planck\ in $S_8$ alone, but a $\sim 2 \sigma$ tension when considering the full multi-dimensional parameter space.
\end{itemize}

\section{Setting up the problem}
\label{sec:simulated}

\begin{figure*}
	\includegraphics[width=0.495\textwidth]{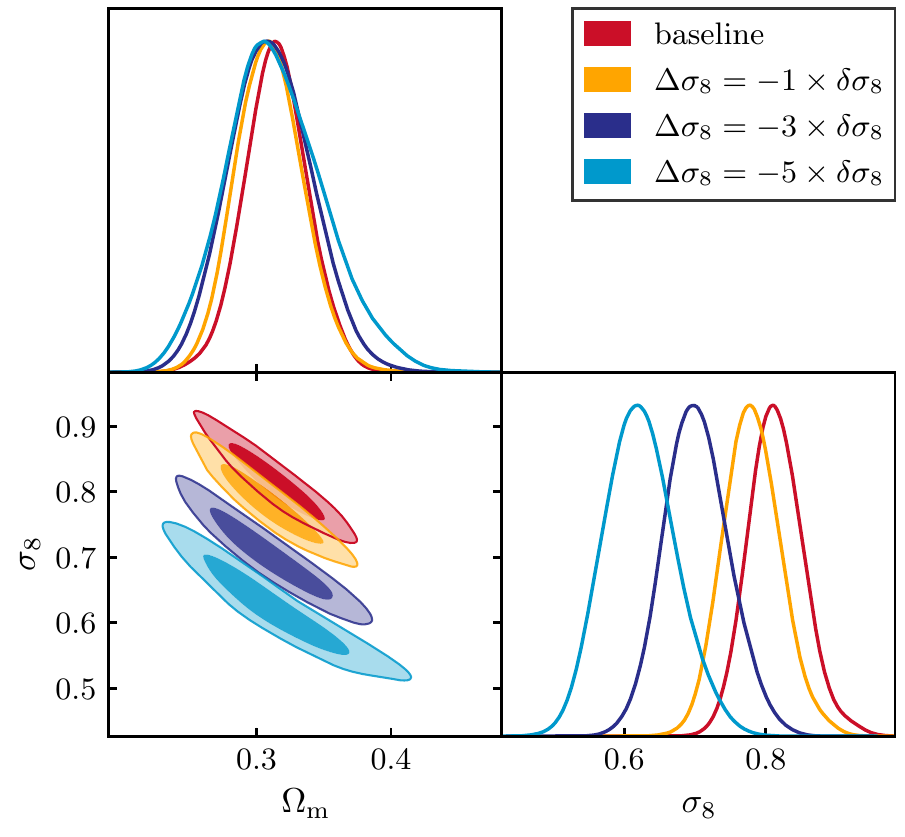}
	\includegraphics[width=0.495\textwidth]{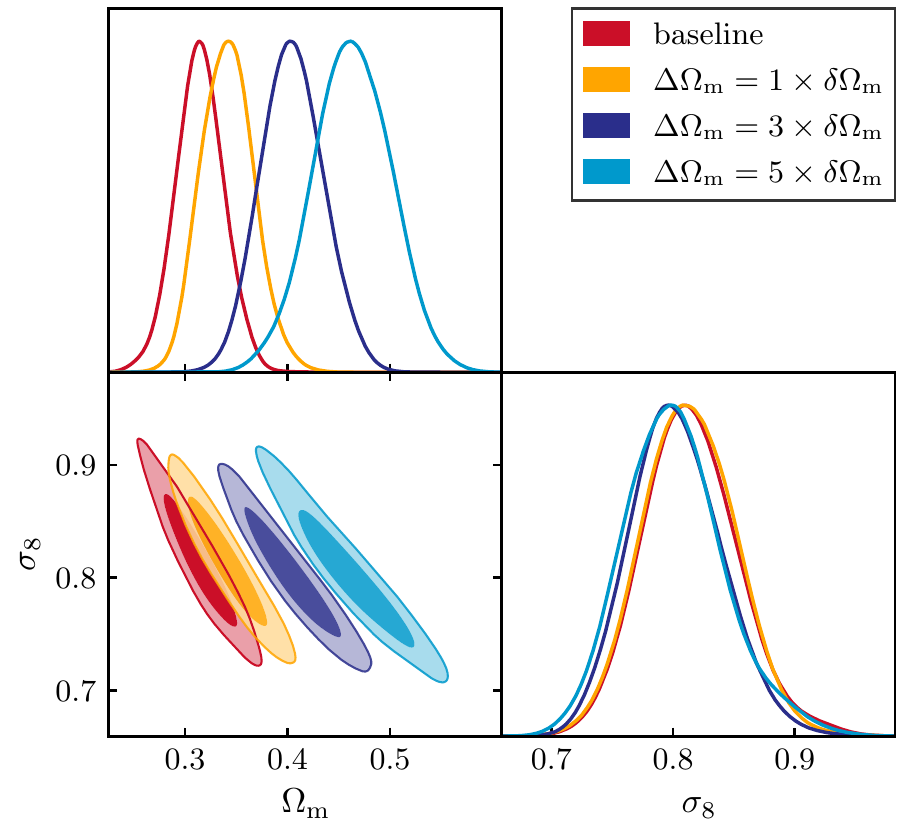}
    \caption{ \label{fig:shifts}
    Marginalized two-dimensional posteriors for some of the simulated DES chains used in this work.
    The darker and lighter shades correspond to the $68\%$ and $95\%$ confidence regions respectively. 
    }
\end{figure*}
The aim of this work is to compare and understand the performance of different metrics for measuring tension between DES and \Planck\ constraints on cosmological parameters. If the two experiments report different values for some cosmological parameters, this might be an indicator that their results are not compatible. However, it is important to understand what this discrepancy means when considering the entire model. 
To do this, we use synthetic DES and \Planck\ data sets that have been generated with different input cosmological parameters in order to produce varying levels of expected tension. By applying the various tension metrics to these synthetic data, we can study how they compare to one another and the known input parameter discrepancies. Note that we 
do not attempt to explain the origin of the possible incompatibility in cosmological parameters reported by two experiments.

We study tension in the context of the flat \LCDM\ cosmological model. Our parameters are
$\left\{ \Omega_{\rm m}, \Omega_{\rm b}, H_0, A_{\rm s}, n_{\rm s} \right\}$,
where $\Omega_{\rm m}$ and $\Omega_{\rm b}$ are the density parameters for matter and baryons, respectively; $H_0$ is the Hubble constant; and $A_{\rm s}$ and $n_{\rm s}$ are respectively the amplitude and slope of the primordial curvature power spectrum 
at a scale of $k=0.05$ Mpc$^{-1}$.
We assume one massive and two massless neutrino species with the total mass equal to the minimum allowed by the oscillation experiments, $m_\nu=0.06$ eV. We do not vary the neutrino mass in our analysis in the simulated data sets, but we do in the reanalysis of tension between DES Y1 and \Planck\ of \cref{sec:y1}, to be consistent with the DES Y1 $3\times 2$pt analysis choices \citep*{DES-MPP:2018}. 
The data and prior choices are further described in \cref{sec:appendix}.

We use the {\tt CosmoSIS} framework\footnote{ \url{https://bitbucket.org/joezuntz/cosmosis/wiki/Home} }~\citep{Zuntz:2014csq} to extract the best-fitting cosmological parameters from the \Planck\ 2015 likelihood by sampling it using Nested Sampling \citep{Skilling:2006}, via the {\tt PolyChord} algorithm\footnote{
\url{https://github.com/PolyChord/PolyChordLite}} \citep{Handley:2015a, Handley:2015b}. From this chain, we infer the best-fit values of the $\Lambda$CDM model parameters according to \Planck\ data and use model predictions from these values to generate a baseline simulated DES-like $3 \times 2$pt data-vector under the \Planck\ cosmology, henceforth referred to as the baseline cosmology. As previously mentioned, the simulated DES data are composed of galaxy clustering, cosmic shear and galaxy--galaxy lensing correlation functions \citep{DES-3x2:2018}. 

\subsection{Generating a-priori tension}

A convenient starting point in our analysis would be synthetically-generated tension in two data sets, corresponding to data vectors generated at different values of cosmological parameters. Precisely how different these two sets of cosmological parameters are should be guided by some preliminary measure of tension. This starting point is henceforth referred to as the "a-priori Gaussian tension", and in this subsection we provide a recipe to define it.

Quantifying the a-priori tension at parameter level with some metrics would make our exercise circular and unfair to other metrics, so it is not a good option. To make progress, we follow a procedure that at least guarantees that the amount of tension we introduce is increasing 
with increasing shifts, and
is, by construction, sensitive to parameters of interest.
Using the \Planck\ and DES posteriors obtained from their respective baseline data vectors, we first compute the variance in the marginalized one-dimensional posterior distributions for $\Omega_{\rm m}$ and $\sigma_8$, referred to as $\var(\theta)$,
where $\theta \in \left\{ \Omega_{\rm m}, \sigma_8 \right\}$. We then shift each parameter by a multiple of the quantity 
\begin{equation} \label{eq:var}
\delta \theta = \sqrt{ {\rm var}(\theta_{\rm DES})+{\rm var}(\theta_{\Planck})}
\end{equation}
and generate simulated DES data vectors with either $\Omega_{\rm m}$ or $\sigma_8$ shifted by integer multiples of the corresponding $\delta\theta$. 
We indicate the total shift with $\Delta\theta \equiv \alpha \delta \theta$ for a given integer $\alpha$.
We then use those data vectors to obtain simulated DES chains.
We shift $\sigma_8$ towards lower values than \Planck's, and $\Omega_{\rm m}$ towards higher values, for simplicity, but we would expect to obtain similar results if the 
shifts were done in the opposite directions.

A shift in $\sigma_8$ is obtained by changing the input value of $A_{\rm s}$. Shifting $\Omega_{\rm m}$, on the other hand, changes the history of structure growth and thereby $\sigma_8$; we compensate for this collateral shift in $\sigma_8$ by counter-shifting $A_{\rm s}$. The DES constraints (shown in the $\Omega_{\rm m}$--$\sigma_8$ plane) from a representative subset of these shifted synthetic data are shown in \cref{fig:shifts}.

If we approximate the difference between the \Planck\ and DES posteriors as a Gaussian distribution in multiple dimensions we can now ask, {\it a priori}, 
what the significance of these shifts is (in the $\Omega_{\rm m}$--$A_{\rm s}$  
plane)
by computing
\begin{equation} \label{Eq:InputTension}
\chi^2 = \delta \theta^T (\mathcal{C}_{\rm D} + \mathcal{C}_{\rm P})^{-1} \delta \theta
\end{equation}
where $\mathcal{C}_{\rm D}$ and $\mathcal{C}_{\rm P}$ are the $2\times 2$ covariance matrices in ($\Omega_{\rm m}, A_{\rm s}$) for DES and \Planck\, respectively. Because we are changing only two parameters, the quantity has two degrees of freedom.
Note that this is just the generalization of Eq.~\eqref{eq:nsigma_1d} to multiple dimensions.
While the Gaussian approximation is not expected to be accurate, especially in the tails of the posteriors, it is expected to be a reasonable guess of the tension that we are inputting into our synthetic examples.

\begin{figure}
	\includegraphics[width=\columnwidth]{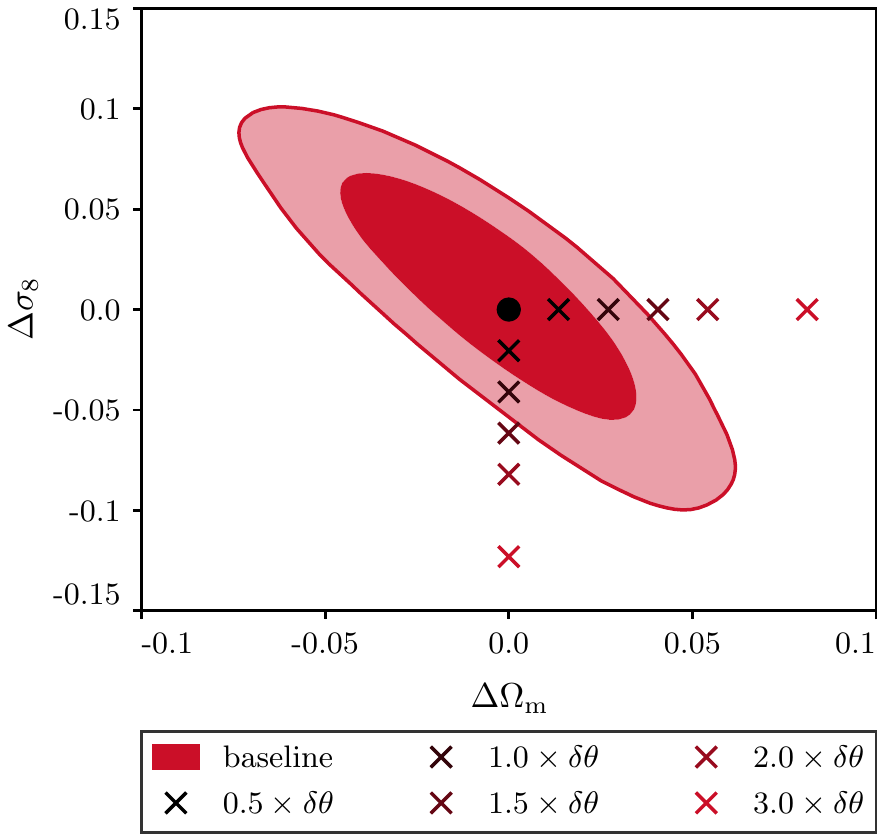}
    \caption{ \label{fig:diff}
    $68 \%$ and $95 \%$ confidence regions of the constraint on the differences in parameters as measured by DES and \Planck, constructed as discussed in \cref{sec:simulated}. The markers indicate the location of the synthetic input shifts. The corresponding a-priori Gaussian tension is shown in \cref{tab:input}.
    }
\end{figure}

\cref{fig:diff} shows the distribution of shifted parameter combinations we describe above, as well as the baseline \Planck\ + DES parameter constraints. Specifically, the contour shows the combined baseline \Planck\ + DES constraints,
while the markers show the best-fit values of individual shifted DES-only constraints.
We can immediately see that, in multiple dimensions, the tension that we attributed to a one-dimensional shift is higher since $\Omega_{\rm m}$ and $\sigma_8$ are correlated.

To quantify the significance of the shifts shown in \cref{fig:diff},
we calculate from Eq.~\eqref{Eq:InputTension} the probability to exceed (PTE) our input shifts in the Gaussian case.
For example, we would like to associate a `one-sigma tension' to an $\Omega_{\rm m}$ shift that lies precisely on the edge of the $68\%$ confidence region. We thus adopt a simple 1D Gaussian conversion
\begin{equation} \label{eq:nsigma}
N_{\sigma} \equiv \sqrt{2}\,{\rm Erf}^{-1}({\rm PTE}),
\end{equation}
where ${\rm Erf}^{-1}$ is the inverse error function. 
Given a probability to exceed, $N_{\sigma}$ matches that probability with the number of standard deviations that an equivalent event from a 1D Gaussian distribution would have. 
Note that the conversion in Eq.~(\ref{eq:nsigma}) is only a convenient proxy to report high statistical significance results, and does not assume Gaussianity {\it per se} in any of the statistics.

The resulting evaluation of the a-priori Gaussian tension is shown in \cref{tab:input}. Here the first column shows the parameter shift applied to DES data in the $(\Omega_{\rm m}, \sigma_8)$ space, where each parameter is shifted by a half-integer multiple of its reported (marginalized) error. The second column shows the full-parameter-space tension calculated using Eq.~(\ref{eq:nsigma}) as described above. 
Note that the `input shifts' in $\Omega_{\rm m}$ lead to higher tension than those in $\sigma_8$. This is because shifting $\Omega_{\rm m}$ while keeping $\sigma_8$ fixed also leads to a shift in $A_{\rm s}$, which increases the tension in the full-dimensional space. 

Finally, let us note that the a-priori tension, by its construction, does not contain stochastic noise, as it effectively measures the distance in the space of input cosmological parameters. This is in contrast with all of the tension metrics that we study below, which are applied to  random realizations of data that do contain noise. The fact that the effectively noiseless input tension is being compared to tension measurements applied on noisy data is one reason why we do not expect a perfect match between the two. We will return to this point in Sec.~\ref{sec:results}.

\begin{table}
\begin{center}
\begin{tabular}{|c|c|}
        \hline \hline
        \multicolumn{2}{c}{Evaluation of a-priori Gaussian tension}
        \\
        \hline\hline
        $(\Omega_{\rm m}, \sigma_8)$ shift & full-par-space $N$-$\sigma$ \\
        \hline 
        $\Delta \sigma_8 = -0.5\,\times \delta \sigma_8$  & $0.02\,\sigma$ \\
        $\Delta \Omega_{\rm m} = +0.5\,\times \delta \Omega_{\rm m}$ & $0.09\,\sigma$ \\
        \hline 
        $\Delta \sigma_8 = -1\,\times \delta \sigma_8$ & $0.4\,\sigma$ \\
        $\Delta \Omega_{\rm m} = +1\,\times \delta \Omega_{\rm m}$ & $1.0\,\sigma$ \\
        \hline
        $\Delta \sigma_8 = -1.5\,\times \delta \sigma_8$ & $1.1\,\sigma$ \\
        $\Delta \Omega_{\rm m} = +1.5\,\times \delta \Omega_{\rm m}$ & $2.3\,\sigma$ \\
        \hline 
        $\Delta \sigma_8 = -2\,\times \delta \sigma_8$ & $2.0\,\sigma$ \\
        $\Delta \Omega_{\rm m} = +2\,\times \delta \Omega_{\rm m}$ & $3.8\,\sigma$ \\
        \hline
        $\Delta \sigma_8 = -3\,\times \delta \sigma_8$ & $3.7\,\sigma$ \\
        $\Delta \Omega_{\rm m} = +3\,\times \delta \Omega_{\rm m}$   & $>5 \,\sigma$ \\
        \hline
        $\Delta \sigma_8 = -5\,\times \delta \sigma_8$ & $>5 \,\sigma$ \\
        $\Delta \Omega_{\rm m} = +5\,\times \delta \Omega_{\rm m}$   & $>5 \,\sigma$ \\
        \hline\hline    
\end{tabular}
\end{center}
\caption{
Evaluation of a-priori Gaussian tension for controlled shifts in ($\sigma_8$ and $\Omega_{\rm m}$). The $\delta\theta$ by whose half-integer value we are shifting these parameters is referring to their respective 1D marginalized posterior as in Eq.~\eqref{eq:var}.
See Eq.~\eqref{eq:nsigma} 
for the explanation how we convert these shifts into the "number of sigmas" in the full parameter space, shown in the second  column. 
}
\label{tab:input}
\end{table}

\section{Tension Metrics}
\label{sec:metrics}

This section describes the tension metrics that we will be comparing in this work. 
Several metrics have been proposed for quantifying tension between cosmological data sets. 
In this work, we select a series of methods that we believe to be appropriate to our data, and which are distinct enough to highlight the strengths and failure modes of each metric. 
We separate the tension metrics into two subcategories, since while all methods aim to quantify tension between data sets, they answer slightly different questions: 

\begin{itemize}
    \item{\bf Evidence-based methods} seek to answer the question:\\ 
    {\it Given hypothesis $H_1$: `The assumed model is capable of generating the data observed by both experiments', and hypothesis $H_2$:  `The assumed model is not capable of generating the data observed by both experiments', which hypothesis is preferred by the data under the assumed model'?} 
    
    \item {\bf Parameter-space methods} seek to answer the question: \\
 {\it What is the statistical 
    significance of the differences between the posteriors for experiments A and B, within the parameter space analyzed by both experiments?}

\end{itemize}

All of the tension metrics that we consider solve the problems that we have discussed in Sec.~\ref{sec:motivation} by considering all dimensions of parameter space. In addition, since they provide results in terms of probabilities, they are independent of the specific parametrizations that are used.

The remainder of this section describes these tension metrics. The results for these metrics will be shown in Sec.~\ref{sec:results}.

\subsection{Bayesian evidence ratio}
\label{sec:bayesr}

\begin{figure*}
	\includegraphics[width=\textwidth]{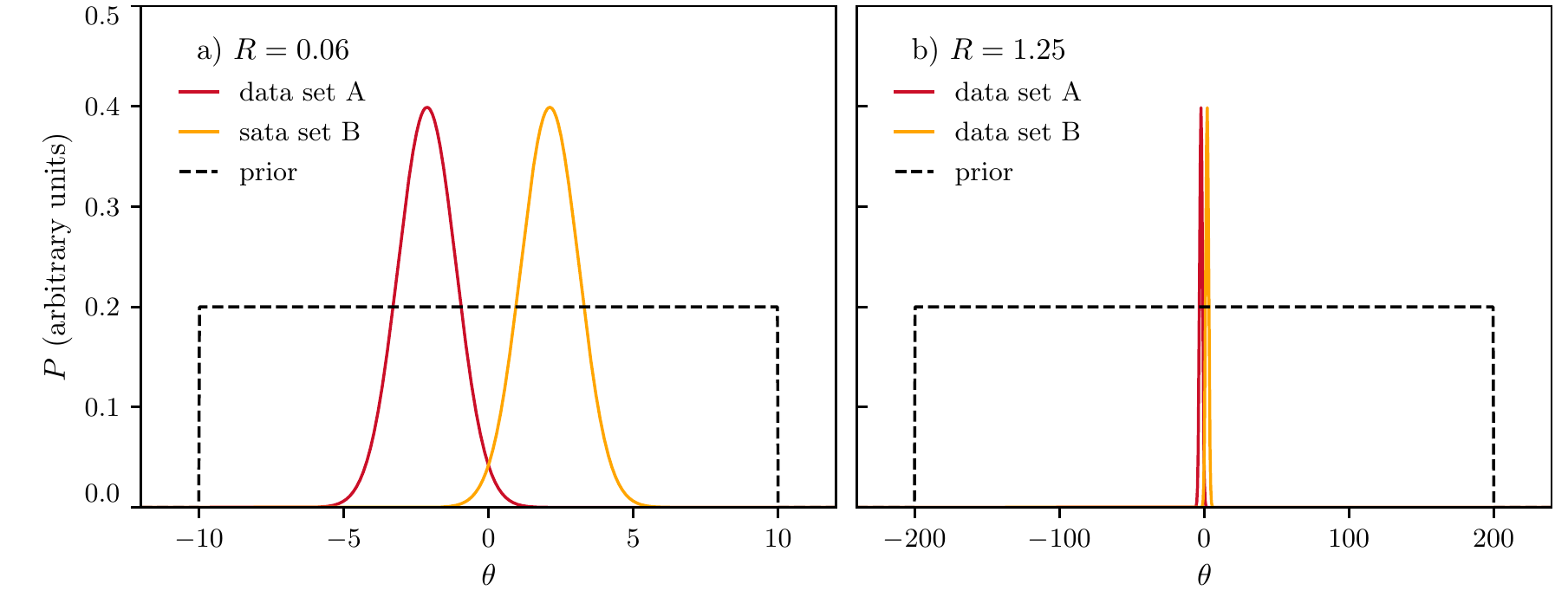}
    \caption{\label{fig:R_priors} 
    Example of the prior-volume dependence of $R$. In amber and red are two gaussians that are at a $3 \sigma$ tension. The black dotted line is the prior (note that it is not normalized, to make it easier to visualize). When we use a uniform prior in the range $[-10, 10]$ (left panel), $R$ is much smaller than one, which means the data sets are in tension. When we increase the prior to $[-200, 200]$ (right panel), $R$ becomes greater than one, indicating agreement. This example, although extreme, illustrates a possible issue of the Bayes ratio as a tension metric.
    }
\end{figure*}

The Bayesian evidence ratio, or Bayes ratio $R$, is an evidence-based method, defined for independent data sets $A$ and $B$ as \citep{Marshall:2004zd}: 
\begin{equation}
\label{eq:bayesr}
    R \equiv \frac{\Z_{AB}}{\Z_A \Z_B}.
\end{equation}
Here, $\Z_D$ is the Bayesian Evidence, defined as the probability of measuring the observed data $D$ for a given model $M$, which can be obtained marginalising over all the model parameters $\theta$:
\begin{equation}
    \Z_D \equiv P(D | M) = \int \rmd \theta \ P(D |\theta, M) P(\theta | M).
\end{equation}
Henceforth, we adopt the following notation for Bayes' theorem:
\begin{equation}
    \P = {\L \times \Pi \over \Z}
\end{equation}
where $\P \equiv P(\theta | D, M)$ is called the posterior, $\L \equiv P(D | \theta,  M)$ is the likelihood, and $\Pi \equiv P(\theta | M)$ is the prior.
The Bayesian Evidence is a difficult quantity to calculate, as it requires integrating a  probability distribution over a large number of dimensions. One of the most frequently-used tools to calculate Bayesian Evidences is Nested Sampling \citep{Skilling:2006}, which also produces posterior distributions. There exist publicly-available codes for Nested Sampling calculations, such as {\tt MULTINEST} \citep{Feroz:2008} and {\tt POLYCHORD} \citep{Handley:2015a, Handley:2015b}. 

In the Bayes ratio $R$ as written in \cref{eq:bayesr}, the numerator requires both data sets to be simultaneously explained by the same parameter values within the model, while the denominator allows each data set to be explained by different parameter values (still within the same assumed underlying model). A more intuitive interpretation  \citep{Amendola:2012wc, Raveri:2018wln, Handley:2019wlz} uses Bayes theorem to rewrite this as 
\begin{equation}
R = \frac{P(A|B,M)}{P(A|M)}     \,,
\end{equation}
(where data sets $A$ and $B$ can be interchanged). That is, does the existence of data set $B$ make the data set $A$ more or less likely than it would be in the absence of $B$, all within the context of assumed model $M$?
Therefore, a ratio of probabilities $R \gg 1$ is interpreted as the data sets being consistent, while $R \ll 1$  indicates that the data sets are in tension. This tension metric has several desirable properties: it is a global statistic (that is, operates on the full parameter space), and it is symmetric between data sets (so tension between data A and data B is the same as tension between B and A). For these reasons, $R$ was used in \cite{DES-3x2:2018}, to quantify tension between the DES Y1 measurements and external data sets. 

\begin{table}
	\centering
	\begin{tabular}{c|c} 
		$\log R$ & Interpretation\\
		\hline
		$>2.3$ & Strong agreement\\
		$(1.2,2.3)$ & Substantial agreement\\
		$(-1.2,1.2)$ & Inconclusive\\
		$(-2.3,-1.2)$ & Substantial tension\\
		$<-2.3$ & Strong tension\\
		\hline
	\end{tabular}
	\caption{Jeffreys' scale used by~\citep{DES-3x2:2018} to quantify agreement or tension between data sets~\citep{Jeffreys:1939}.
	}
	\label{tab:jeffreys}
\end{table}

This new interpretation carries an important issue, which is $R$'s dependence on the prior volume: as described by \cite{Handley:2019wlz}, \cref{eq:bayesr} can be rewritten as: 
\begin{equation}
    R \equiv \int \rmd \theta \ \frac{\P_A \P_B}{\Pi}.
\end{equation}
For a flat and uninformative prior, 
$R$ is therefore proportional to the prior volume. For example, doubling the prior volume doubles 
the value of $R$, 
and increases the agreement between the data sets independently of the shape of the posteriors. As an extreme case, one could increase the prior range arbitrarily to make any two posteriors consistent according to $R$. This is illustrated by \cref{fig:R_priors}, which gives two equal-width Gaussians horizontally offset by $3 \sigma$. The Bayes ratio is close to zero when the prior encompasses relatively tightly the bulk of the two distributions, but goes up to $R > 1$ if the prior is made sufficiently wide. 
In the latter case, the Bayes-ratio-logic says that the two Gaussians are close to each other \textit{relative to the width of the prior}, and hence are reported to not be in any tension. This prior dependence is therefore a central feature of the Bayes ratio. Nevertheless, such a prominent role for the prior may be worrying in situations when physically-motivated priors are not available. 

A second concern 
about the Bayes ratio $R$ is that its raw numerical value needs calibration. $R$ is the ratio of probabilities (see \cref{eq:bayesr}) and one often uses the Jeffreys' scale (\cite{Jeffreys:1939}; see \cref{tab:jeffreys}) to convert the different outcomes to interpretations about the presence of tension between data sets. However, the boundaries in Jeffreys' scale are arbitrary, and they lack obvious interpretation as a statistical significance. 

Both the interpretation and the calibration problem can be circumvented if another tension metric is used to calibrate the Bayes ratio. In this paper, we use the simulated data vectors described in \cref{sec:simulated} to calibrate the Bayes ratio outcomes (along with those from other tension metrics). Note, however, that this calibration is very specific to our choice of the problem, such as the observables, the parameter space, or the priors we employ. Our results would not be generalizable to an arbitrary cosmological analysis.

\subsection{Bayesian Suspiciousness}

Bayesian Suspiciousness \citep{Handley:2019b} is an evidence-based method, introduced as an alternative to the Bayes ratio from \cref{sec:bayesr} for the case of priors which, instead of being motivated by prior knowledge, are purposefully wide and uninformative. This is the case for DES, where wide priors are chosen with the goal of obtaining DES-only constraints. The idea is the following: We divide the Bayes ratio $R$ in two parts, one that quantifies the probability of the data sets matching given the prior width, and another one that quantifies their actual mismatch. The first part is quantified by the information ratio $I$, defined as: 
\begin{equation}
    \log I \equiv \D_A + \D_B - \D_{AB},
\end{equation}
where $\D$ is the Kullback--Leibler Divergence \citep{Kullback:1951}:
\begin{equation}
    \D \equiv \int \P \log \left( {\P \over \Pi} \right) \ \rmd \theta \,.
\end{equation}
The Kullback--Leibler Divergence is particularly well suited to eliminate the prior dependence from the Bayes ratio, as it quantifies how much information has been gained going from the prior $\Pi$ to the posterior $\P$. Therefore, it encloses the prior dependence that we want to eliminate. The Kullback--Leibler Divergence has been extensively used in cosmology \citep[e.g.][]{Hosoya:2004, Verde:2013, Seehars:2014, Seerhars:2016, Grandis:2016, Nicola:2019}.

The part of the Bayes ratio $R$ that is left after subtracting the dependence on prior volume depends only on the actual mismatch between the posteriors, and it is what we call Bayesian Suspiciousness $S$: 
\begin{equation}
    \log S = \log R - \log I.
\end{equation}
As explained in \cref{sec:bayesr} and in \cite{Handley:2019wlz}, the main concern regarding the Bayes ratio $R$ is that the tension can be `hidden' by widening the priors. $S$ can be understood as the version of $R$ that corresponds to the smallest priors that do not significantly alter the posterior. 
It also has two useful qualities that $R$ lacks: It does not depend on the prior volume and, in the case of Gaussian posteriors, it follows a $\chi^2_d$ distribution, where $d$ is the effective number of degrees of freedom constrained by both data sets. Therefore, we can assign a {\it tension probability} $p_T$ as the p-value of the distribution. This tension probability quantifies the probability of the observed tension occurring by chance. While the chi-squared interpretation relies on the approximation of Gaussian posteriors,\footnote{As pointed out by \cite{Handley:2019wlz}, non-Gaussian posteriors can be `Gaussianized' using Box--Cox transformations \citep{Box:1964, Joachimi:2011, Schuhmann:2016}, that preserve the value of $S$. Therefore, the chi-squared interpretation of $S$ derived in the Gaussian case can be approximately valid even for posteriors that do not look Gaussian, even if it is not guaranteed that both posteriors can be Gaussianized simultaneously.} the rest of this section does not, so the value and sign of $S$ can be used to measure tension for any posterior distributions. 

To obtain the value of $p_T$, we need to calculate the effective number of dimensions constrained by the combination of the data sets. While there are several available methods to do this, we propose using the Bayesian Model Dimensionality \citep{Handley:2019b}:
\begin{equation}
\label{eq:BMD}
    d = 2 \int \P \left( \log {\P \over \Pi} - \D \right)^2. 
\end{equation}
This formula is analogous to the more traditional Bayesian Model Complexity (BMC) \citep{Spiegelhalter:2002} used in previous cosmological analyses \citep[e.g.][]{Kunz:2006, Bridges:2009}, with which it shares the property that it is formed of Bayesian quantities and recovers a value of $d=1$ for the 1D Gaussian case. But while the BMC requires the use of either the mean or maximum-posterior parameter values and is hence subject to sampling error (i.e.\ numerical noise due to a finite length of an MCMC chain), \cref{eq:BMD} does not suffer from these issues \citep{Handley:2019b}. 

While the Suspiciousness is according to our definition an evidence-based method, it has been 
recently shown~\citep{Heymans:2020} that it can be reformulated as the difference of the log-likelihood expectation values of joint and individual data sets, leading to a relation between the suspiciousness and the goodness-of-fit loss introduced in~\cref{Sec:GoFloss}~\citep{Joudaki:2020shz} through the Deviance Information Criterion~\citep{Spiegelhalter01bayesianmeasures} 
This shows
that despite them being defined very differently, there are fundamental relations between these statistics.

All the quantities discussed in this subsection can be simply obtained from a single nested sampling chain (in the case of the BMD, or even an MCMC chain), which means that their computational cost is the same as that of the Bayes ratio introduced in \cref{sec:bayesr}. Nested sampling can also give us an estimate 
of the sampling error, by re-sampling the sample weights \citep{Higson2018}. \cite{Joachimi:2020abi} noted that this method can lead to noise in the dimensionality calculation. This noise was included in this work, and contributes to the error in the estimate of the tension probability. All calculations are implemented in the python package {\tt anesthetic}\footnote{
\url{https://github.com/williamjameshandley/anesthetic}
}
\citep{anesthetic};
an example on how to calculate these quantities can be found at \url{https://github.com/Pablo-Lemos/Suspiciousness-CosmoSIS}.

\subsection{Parameter differences}
Another estimator that we consider is the Monte Carlo estimate of the probability of a parameter difference as described in~\cite{Raveri:2019gdp}. This is a parameter-space method, which relies on the computation of the parameter difference probability density $\P(\Delta \theta)$.
In the case of two uncorrelated data sets this is given by the convolution integral:
\begin{equation} \label{Eq:ParameterDifferencePDF}
\P(\Delta \theta) = \int_{V_p} \P_A(\theta) \P_B(\theta-\Delta \theta) d\theta
\end{equation}
where $P_A$ and $P_B$ are the two parameter posterior distributions and $V_p$ is the support of the prior, i.e. the region of parameter space where the prior is non-vanishing.
Notice that this probability density has been marginalized over the value of the parameters and only constrains their difference.

Once the density of parameter shifts is obtained one can quantify the probability that a genuine shift exists: 
\begin{equation} \label{Eq:ParamShiftProbability}
\Delta = \int_{\P(\Delta\theta)>\P(0)} \P(\Delta\theta) \, d\Delta\theta
\end{equation}
which is the posterior mass above the iso-probability contour for no shift, $\Delta\theta=0$.
Note that since Eq.~\eqref{Eq:ParamShiftProbability} is the integral of a probability density it is invariant under reparametrizations.

Equations~(\ref{Eq:ParameterDifferencePDF}) and (\ref{Eq:ParamShiftProbability}) look straightforward, but their evaluation is greatly complicated in parameter spaces with a large number of dimensions. In such cases (which are typical in cosmological applications), the posterior samples cannot be easily smoothed or interpolated to a continuous function, and we are left to work exclusively with $N_A$ samples from the posterior $P_A$ and $N_B$ from $P_B$, i.e. discrete representations of the posteriors of interest. Each one of the $N_AN_B$ pairs of samples corresponds to one term on the right-hand side of Eq.~(\ref{Eq:ParameterDifferencePDF}) (with $\Delta\theta=\theta_A-\theta_B$, where $\theta_A$ and $\theta_B$ are the parameter values for that pair).\footnote{In the case of weighted samples the weight of the parameter difference sample is the product of the two weights.} 

To make progress, we perform the integral in Eq.~(\ref{Eq:ParamShiftProbability}) with a Monte Carlo algorithm. 
One computes the Kernel Density Estimate (KDE) probability of $\Delta\theta=0$ and then the KDE probability of each of the samples of the parameter difference posterior.
The number of samples with KDE probability above zero divided by the total number of samples is the Monte Carlo estimate of the integral in Eq.~(\ref{Eq:ParamShiftProbability}) and the error can be estimated from the binomial distribution.
This approach largely mitigates the need for an accurate estimate of the optimal KDE smoothing scale. 
In practice we use a multivariate Gaussian kernel with smoothing scale fixed by the Silverman's rule~\citep{chacon2018multivariate}.

We use the implementation of this tension estimator in the {\tt tensiometer}\footnote{\url{https://github.com/mraveri/tensiometer}} code.

\subsection{Parameter differences in update form} \label{Sec:ParameterUpdate}
Another parameter-space method that we consider is the update difference-in-mean (UDM) statistic, as defined in~\cite{Raveri:2018wln}. 
This compares the mean parameters determined from one data set, $\hat{\theta}^{A}$, with their updated value, $\hat{\theta}^{A+B}$, obtained after adding another data set.
The shifts in parameters are then weighted by their inverse covariance to give
\begin{equation}
Q_{\rm UDM} = (\hat{\theta}^{A+B}-\hat{\theta}^{A})^T \left( \mathcal{C}^A -\mathcal{C}^{A+B} \right)^{-1}(\hat{\theta}^{A+B}-\hat{\theta}^{A})
\end{equation}
where $\mathcal{C}^A$ and $\mathcal{C}^{A+B}$ are the posterior covariances of the single data set $A$ and the joint data set $A+B$. 
If the parameters $\hat{\theta}^{A}$ and $\hat{\theta}^{A+B}$ are Gaussian distributed then $Q_{\rm UDM}$ is chi-squared distributed with ${\rm rank}(\mathcal{C}^A -\mathcal{C}^{A+B})$ degrees of freedom.
These degrees of freedom are the parameters that are measured by both data sets $A$ and $B$ and are the only ones that can actively contribute to a tension between the two.
For both fully informative and uninformative priors the statistical significance of a shift in $\hat{\theta}^{A+B}-\hat{\theta}^{A}$ is the same as the shift in 
$\hat{\theta}^{A}-\hat{\theta}^{B}$ since both of them are weighted by their inverse covariance. 
We note that in non-update form and for uninformative priors, i.e. Eq.~\eqref{Eq:InputTension}, parameter differences are equivalent to the Index of Inconsistency~\citep{Lin:2017bhs, Lin:2017ikq, Lin:2019zdn}, while providing a clear assessment of statistical significance rather than interpretation on the Jeffreys' scale.

There are two main advantages of using $Q_{\rm UDM}$ instead of non-update difference in mean statistics:
parameter-space directions that can exhibit interesting tension are identified {\it a priori}, i.e.\ before explicitly measuring the tension, to aid physical interpretation; non-Gaussianities are mitigated since we can select the most constraining and Gaussian of two data sets.

As shown in~\cite{Raveri:2018wln}, an effective method to compute $Q_{\rm UDM}$ in practice consists of breaking down the calculation as a sum over the Karhunen--Lo\'eve (KL) modes of the covariances involved.
We indicate these modes with $\phi^a$ and their corresponding
generalized eigenvalue with $\lambda^a$. The modes $\phi^a$ are uncorrelated for both data set $A$ and $A+B$. 
For a given KL mode $\lambda^a-1$ is the improvement observed for the variance in the value of that mode when the second data set is added to the first.
To avoid sampling noise in the calculation of $Q_{\rm UDM}$ we restrict our calculation to modes that satisfy:
\begin{equation}
0.2 < \lambda^a-1 < 100 \,.
\end{equation}
The lower bound removes directions along which data set B is not updating A, while the upper bound removes directions along which A is not updating B.
In both cases, with perfect knowledge of the covariances these directions would not contribute to the end result.

We notice here that the procedure of identifying the KL modes can be performed {\it a priori}, before looking at the data, starting from the Fisher matrix.
We also point out that the set of KL modes is invariant under linear parameter transformations while the principal-component decomposition is not.

The KL decomposition of parameter shifts allows to investigate the physical origin of the reported tensions.
As discussed in~\cite{Wu:2020nxz} we can write the parameters' Fisher matrix $F=(\mathcal{C})^{-1}$ as a sum over KL components:
\begin{equation}
F_{\alpha\alpha} = \sum_{a}F_{\alpha\alpha}^a = \sum_a \phi^a_\alpha\phi^a_\alpha/\lambda^a \,.
\end{equation}
The fractional Fisher information $F_{\alpha\alpha}^a/F_{\alpha\alpha} \in [0,1]$ tells us how important a given KL mode is in constraining a cosmological parameter. 
Low values mean that the KL mode can be removed from the full decomposition without altering the parameter constraint.

\begin{figure}
\includegraphics[width=\columnwidth]{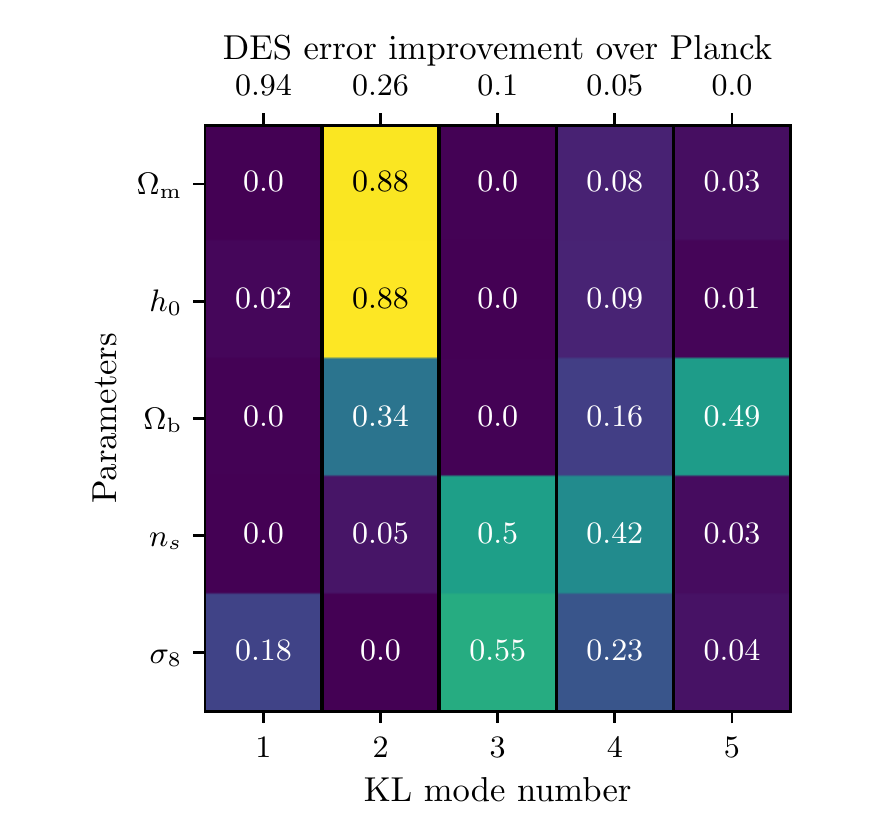}
\caption{ \label{fig:kl}
The fractional Fisher information on cosmological parameters for \Planck\ computed using the KL modes from its update with simulated DES.
Each line shows the fractional contribution of each KL mode to the total information on a given parameter. The sum of values in each row is one.
The numbers on top of the figure show the fractional error improvement of DES over Planck for each KL mode.
}
\end{figure}

In \cref{fig:kl} 
we show the fractional contribution of different KL modes to the {\it Planck} Fisher matrix when it is updated with our simulated DES measurements.
We also report in the figure the error improvement which is given by $\sqrt{\lambda^a}-1$ for each mode.
We have a total of five modes, equal to the number of parameters that the data sets have in common and we have sorted them by error improvement of DES+\Planck\ over \Planck\ alone.
The first data set -- in this case \Planck\ -- is setting the parameter combinations that are updated for each mode, while the second data set is setting the improvement factor.
For the first two modes we can see that DES improves on the \Planck\ determination of $\sigma_8$ by almost a factor two (94\%) and the determination of $\Omega_{\rm m} h^2$ by 26\%. DES does not improve other modes significantly.

We use the implementation of $Q_{\rm UDM}$ and related KL decomposition algorithms in the {\tt tensiometer} code.

\subsection{Goodness-of-fit loss} \label{Sec:GoFloss}
We next consider Goodness-of-fit loss which measures how much goodness-of-fit degrades when joining two data sets. 
This is a method in between evidence- and parameter-based ones since it relies on both likelihood values and parameters. 
When fitting two data sets separately, each probe can individually invest all model parameters in improving its goodness of fit. However,
when the two measurements are joined the parameters have to compromise and the quality of the joint fit naturally degrades.
This degradation is quantified by the estimator:
\begin{equation} \label{Eq:QDMAP}
Q_{\rm DMAP} = 2\ln \mathcal{L}_A(\theta_{pA}) +2\ln \mathcal{L}_B(\theta_{pB}) -2\ln \mathcal{L}_{A+B}(\theta_{pA+B})
\end{equation}
where $\theta_{pA}, \theta_{pB}$ and $\theta_{pA+B}$ are the Maximum a Posteriori (MAP) parameters measured by the first and second probe and their combination respectively, and $\mathcal{L}$ is the data likelihood for the single and joint probes and is evaluated at the Maximum a Posteriori (MAP) point, $\theta_{p}$.  
We use the subscript DMAP to denote the difference in MAP estimates. 
As discussed in~\cite{Raveri:2018wln}, when the likelihoods and posteriors are Gaussian $Q_{\rm DMAP}$ is $\chi^{2}$ distributed with
\begin{equation} \label{Eq:DeltaNeff}
\Delta N_{\rm eff} = N_{\rm eff}^A +N_{\rm eff}^B -N_{\rm eff}^{A+B}
\end{equation}
degrees of freedom where $N_{\rm eff}^A$, $N_{\rm eff}^B$, and $N_{\rm eff}^{A+B}$ are the respective numbers of the degrees of freedom
\begin{equation} \label{Eq:NumberEffectiveParameters}
N_{\rm eff} = N -{\rm tr}[\mathcal{C}_\Pi^{-1}\mathcal{C}_p]
\end{equation}
is the number of parameters that a data set ends up constraining compared to the priors it began with.
The goodness-of-fit is expected to degrade by one for each measured parameter, and indicates tension if the decrease is higher.
Only the parameters that are constrained by the data over the prior can contribute to a tension since prior-constrained parameters cannot be optimized to improve the data fit. 
In the limits where the prior is uninformative or fully informative $Q_{\rm DMAP}$ is the likelihood expression for parameter shifts discussed in the previous sections and its statistical significance should match the one obtained with parameter-shift techniques.

Notice that this estimator requires Gaussianity in both data space and parameter space. This is a stronger requirement than just approximate Gaussianity in parameter space, 
and limits its applicability in practice.
Most of the likelihoods that we use here are Gaussian in data space with the exception of the large-scale CMB likelihood.
This can be thought to be a prior on the optical depth of reionization, $\tau$, that would not contribute to the tension budget since it is not shared with DES and hence allows us to use $Q_{\rm DMAP}$.

We use the implementation of $Q_{\rm DMAP}$ in the {\tt tensiometer} code.

\subsection{Eigentension}\label{sec:eigentension}

The goal of the Eigentension parameter-space method is to identify well-measured eigenmodes in the data and compare the parameter constraints of two experiments within the subspace spanned by the well-measured eigenmodes. Here, we briefly describe the steps taken to quantify the tension between the fiducial \Planck\ and DES constraints in this paper, and refer the reader to \cite{ParkRozo:2019} for a more detailed discussion and testing of the method.

We begin by identifying the well-measured parameter subspace by following these steps:
\begin{enumerate}
\item Obtain the parameter covariance matrix from a set of fiducial constraints for DES and identify the eigenvectors of this covariance matrix.
\item For each eigenvector, take the ratio of its variance in the prior to its variance in the posterior. If this ratio is above $10^{2}$, identify the eigenvector as well-measured or robust.
\item Project the fiducial \Planck\ constraints and the various DES constraints along the subspace spanned by the robust eigenvector(s), and create importance sampled chains of equal length for each constraint.
\end{enumerate}
For (i), we use constraints from a fiducial DES analysis with a noiseless data vector generated from theory under the \Planck\ best-fit parameters and the true DES Y1 covariance matrix. This allows the {\it ad hoc} choice of $10^2$ as the threshold value in (ii), which we make after examining the eigenvectors from (i), to be {\it a priori}. We identify one well-measured DES eigenvector:
\begin{equation}
e_\mathrm{DES} = \sigma_8 \Omega_\mathrm{m}^{0.57}
\end{equation}
that has a variance ratio of 2665, and construct importance sampled chains of length $10^5$ along this eigenmode. With the projected chains in hand, we quantify tension between two constraints $i$ and $j$ as following; we
\begin{enumerate}
\item construct the chain of differences $\Delta e = e_i - e_j$ between the importance sampled chains for $i$ and $j$.
\item approximate the probability surface for $\Delta e$ via KDE
, and identify the iso-probability contour that crosses the origin, i.e. $\Delta e = 0^N$, where $N$ is the number of robust eigenvectors identified.
\item integrate the probability surface within the origin-crossing contour, and convert the integral to Gaussian sigmas.
\end{enumerate}
For (ii), we use a Gaussian KDE with bandwidths determined from Silverman's rule of thumb, and a straightforward Monte Carlo integration with $1.28\times10^7$ random draws,which is  sufficient to quantify 
tensions up to $5.4\sigma$. 

\begin{table*}
\centering
\setlength{\tabcolsep}{0.6em}
\begin{tabular}{|l|ccccccc}
\hline
    \multirow{2}{*}{1D shift} & a-priori  & \multicolumn{2}{c}{Bayes ratio~~} & \multirow{2}{*}{Eigentension} & \multirow{2}{*}{GoF Loss} & MCMC/Update & \multirow{2}{*}{Suspiciousness} \\
     & Tension & $\hfill\log R\hfill$ & Interpretation & & & Param Diffs &  \\   
\hline \hline
     Baseline & $0 \,\sigma$ & $5.7 \pm 0.6$ & Strong Agreement & $0.5\,\sigma$ & $0.2\,\sigma$ & $0.3/0.3\,\sigma$ & $(0.1 \pm 0.1) \,\sigma$ \\
\hline 
    $\Delta \sigma_8 = - 0.5 \times \delta \sigma_8$ & $0.0 \,\sigma$ & $6.4 \pm 0.6$ & Strong Agreement & $0.4\,\sigma$ & $0.4\,\sigma$ & $0.3/0.4\,\sigma$ & $(0.2 \pm 0.2) \,\sigma$ \\
    $\Delta \Omega_{\rm m} = 0.5 \times \delta \Omega_{\rm m}$ & $0.1 \,\sigma$ &$5.4 \pm 0.6$ & Strong Agreement & $1.3\,\sigma$ & $0.7\,\sigma$ & $0.9/0.8\,\sigma$ & $(0.5 \pm 0.2) \,\sigma$ \\
\hline
    $\Delta \sigma_8 = - 1 \times \delta \sigma_8$ & $0.4 \,\sigma$ &$5.5 \pm 0.6$ & Strong Agreement & $1.1\,\sigma$& $0.8\,\sigma$ & $1.0/0.8\,\sigma$ & $(0.3 \pm 0.2) \,\sigma$ \\
    $\Delta \Omega_{\rm m} = 1 \times \delta \Omega_{\rm m}$ & $1.0 \,\sigma$ &$3.5 \pm 0.5$ & Strong Agreement & $2.3\,\sigma$& $1.9\,\sigma$ & $1.8/1.7\,\sigma$ & $(1.5 \pm 0.3) \,\sigma$ \\
\hline
    $\Delta \sigma_8 = - 1.5 \times \delta \sigma_8$ & $1.1 \,\sigma$ & $3.6 \pm 0.6$ & Strong Agreement & $2.0\,\sigma$ & $1.2\,\sigma$ & $1.8/1.9\,\sigma$ & $(1.5 \pm 0.3) \,\sigma$ \\
    $\Delta \Omega_{\rm m} = 1.5 \times \delta \Omega_{\rm m}$ & $2.3 \,\sigma$ & $-0.4 \pm 0.6$ & No Evidence & $3.3\,\sigma$ & $3.0\,\sigma$ & $2.8/2.7\,\sigma$ & $(2.9 \pm 0.4) \,\sigma$ \\
\hline
    $\Delta \sigma_8 = - 2 \times \delta \sigma_8$ & $2.0 \,\sigma$ & $0.3 \pm 0.6$ & No Evidence & $2.6\,\sigma$ & $2.1\,\sigma$ & $2.7/3.0\,\sigma$ & $(2.2 \pm 0.4) \,\sigma$ \\
    $\Delta \Omega_{\rm m} = 2 \times \delta \Omega_{\rm m}$ & $3.8 \,\sigma$ &$-4.8 \pm 0.6$ & Strong Tension & $4.1\,\sigma$ & $3.9\,\sigma$ & $3.4/3.6\,\sigma$ &  $(4.1 \pm 0.6) \,\sigma$ \\
\hline
    $\Delta \sigma_8 = -3 \times \delta \sigma_8$ & $3.7 \,\sigma$ & $-6.2 \pm 0.6$ & Strong Tension & $4.3\,\sigma$ & $3.4\,\sigma$ & $4.6/4.8\,\sigma$ & $(3.7 \pm 0.5) \,\sigma$ \\
    $\Delta \Omega_{\rm m} = 3 \times \delta \Omega_{\rm m}$ & $> 5 \,\sigma$ & $-16.2 \pm 0.6$ & Strong Tension & $>5.4\,\sigma$& $6.2\,\sigma$ & $5.3/5.3\,\sigma$ & $(5.9 \pm 0.7) \,\sigma$ \\
\hline
    $\Delta \sigma_8 = -5 \times \delta \sigma_8$ & $>5 \,\sigma$ & $-26.3 \pm 0.6$ & Strong Tension & $>5.4\,\sigma$ & $5.8\,\sigma$ & $6.8/8.8\,\sigma$ & $(6.3 \pm 0.8) \,\sigma$  \\
    $\Delta \Omega_{\rm m} = 5 \times \delta \Omega_{\rm m}$ & $>5 \,\sigma$  & $-47.0 \pm 0.6$ & Strong Tension & $>5.4\,\sigma$ & $10.0\,\sigma$ & $6.6/8.1\,\sigma$ & $(9.6 \pm 1.2) \,\sigma$ \\
\hline
\end{tabular}
\caption{\label{tab:results}
The tension between \Planck\ and simulated DES chains for different shifts in $\sigma_8$ and $\Omega_{\rm m}$, calculated via the different tension metrics described in the main text. The first column refers to the number of one-dimensional standard deviations by which each parameter is shifted, defined in \cref{eq:var}. The a-priori Gaussian tension is calculated as 
described in \cref{sec:simulated} and serves only as an order of magnitude approximation of expected results.
The probability results of each of the tension metrics is converted to a number of effective sigmas using Eq.~\eqref{eq:nsigma}.
}
\end{table*}

\subsection{Other metrics}

As mentioned in the introductions, a plethora of methods to quantify tension 
can be found in the cosmological literature.
Our work does not investigate all of these methods, as this would make the analysis too wide in scope. For example, 
Hyperparameters~\citep{Hobson:2002zf, Bernal:2018cxc} are more useful to construct a posterior 
from data sets in tension, by factoring in possible unknown systematic effects. The surprise~\citep{Seehars:2015qza} is best suited for experiments that are an update from a previous version with less data. Posterior 
Predictive Distributions~\citep{Feeney:2018mkj} are similar in nature to the Evidence Ratio as shown in \citet{Lemos:2019txn}. Other methods are 
not considered as they closely resemble others, such as \citet{Amendola:2012wc, Martin:2014lra, Joudaki:2016mvz} being based on the Bayesian Evidence ratio, and \citet{Lin:2017ikq, Adhikari:2018wnk, Lin:2019zdn} being different versions of 
parameter differences in update form. 

\section{Results using simulated DES data} \label{sec:results}

\begin{figure*}
	\includegraphics[width=0.9\textwidth]{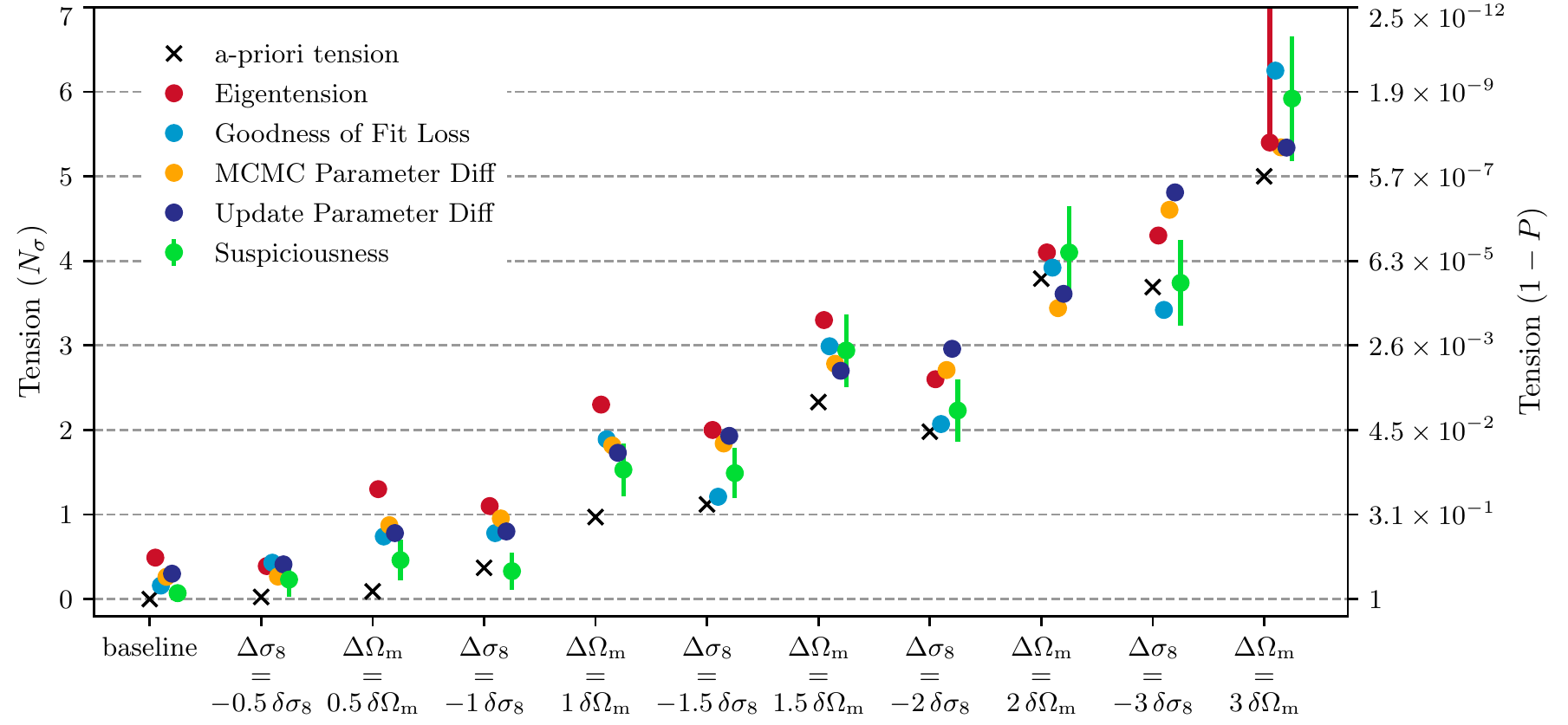}
    \caption{\label{fig:results}
    A graphical illustration of the main results of \cref{tab:results}. 
    Different points show the tension calculated by each tension metric as a function of the input shifts. The error bars in the green points correspond to sampling errors, which can be calculated
    for evidence-based methods by re-sampling the nested sampling weights.
}
\end{figure*}

\begin{figure*}
	\includegraphics[width=0.9\textwidth]{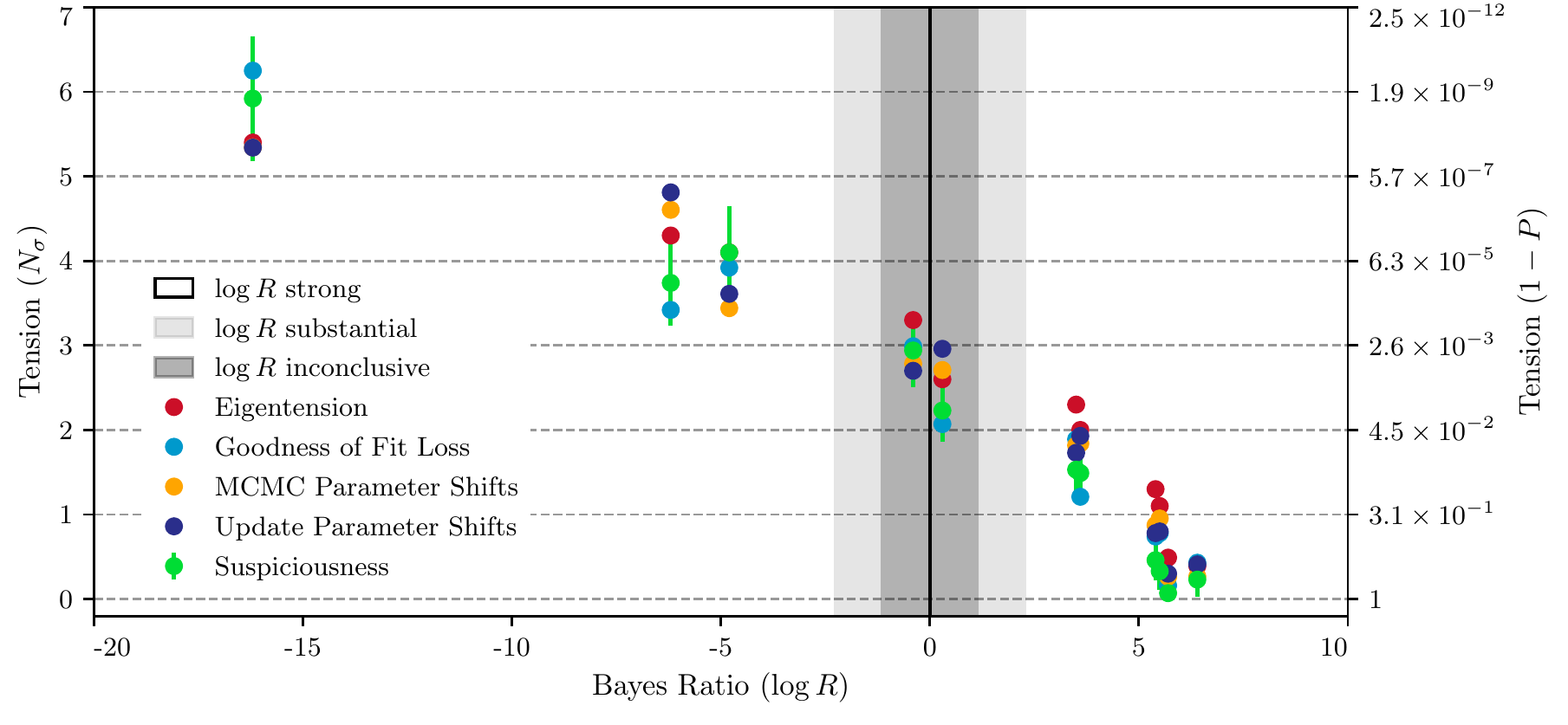}
    \caption{\label{fig:BayesCalibration}
    Tension estimates given by different metrics versus the corresponding Bayes ratio. Shaded regions highlight Jeffreys' scale used to interpret the   Bayes ratio, with the vertical line separating ``Tension" to the left and ``Agreement" to the right.
    }
\end{figure*}

In this section, we apply the tension metrics described in \cref{sec:metrics} to the simulated vectors obtained as outlined in \cref{sec:simulated}, and compare the results to our a-priori expectation from~\cref{sec:simulated}.
Our results are shown in \cref{tab:results} and graphically illustrated in \cref{fig:results}.  

We first note that our estimates of a-priori Gaussian tension should be only used as an rough indication and are generally lower than the tension evaluated by the metrics that we study.
This is because the a-priori Gaussian tension does not have noise in the data vector while the tensions simulations do. This noise realization 
is the same for all the shifts, which explains the fact that the a-priori tension is systematically lower in all results with respect to other
tension estimators. 
We can see this in the baseline case, where in a noiseless case all metrics would obtain perfect agreement (a `$0 \sigma$' tension), but instead the noise leads to small discrepancies.

When applying parameter-shift estimators in both MCMC and update form we can see, from \cref{tab:results} and \cref{fig:results}, that, for tensions measured up to $5\sigma$, the two estimates agree very well, to within $0.3\sigma$. 
This overall result is reassuring since these two estimators are measuring the same sense of tension between the two data sets.
This agreement is also expected since the distributions that we consider are roughly Gaussian in the bulk of the distribution.
At high statistical significance MCMC results are lower in both cases and this suggests that the decay of the tails of the distribution is slower than a Gaussian distribution.
For the parameter update we observe that the two parameter combinations, discussed in \cref{Sec:ParameterUpdate}, that DES$+$\Planck\ significantly improves over \Planck-only do not appreciably change throughout the test cases. 

In case of either fully informative or uninformative priors the statistical significance of GoF loss is expected to match the one reported by parameter-shift estimators.
As we can see from \cref{tab:results} that is the case at low statistical significance.
Non-Gaussianities in the form of slowly-decaying tails violate the assumptions used by the GoF loss estimator, while their impact can be mitigated by parameter shifts in update form. As a result, as statistical significance increases, in \cref{tab:results} the two estimates differ.
In particular, as expected, GoF loss overestimates statistical significance since this estimator is assuming Gaussian decay in the tails.

For Eigentension, we make use of the metric on the simulated vectors, making use of the robust DES eigenvector and the Monte Carlo sampling procedure discussed in Sec.~\ref{sec:eigentension}. Note that the Eigentension metrics are calculated only up to 5.4$\sigma$, or 1 in $1.28\times10^7$; beyond this probability we simply quote that the tension is greater than $5.4\sigma$ and consider the tension to be definitive. The results are in good agreement with other tension metrics, in particular the two parameter shift estimators, with which Eigentension shares the general approach of quantifying tensions at the parameter space level.

With Suspiciousness, as shown in \cref{tab:results} and in \cref{fig:results}, we obtain good agreement with the rest of tension metrics, especially when we consider the sampling error estimated from repeated re-samplings for the weights of the chain. 
To assign a tension probability, we need to calculate the Bayesian Model Dimensionality, for which we get $d = 2.3 \pm 0.1$. 
At high statistical significance, Suspiciousness seems to agree particularly well with GoF loss. This is reassuring since the two estimators coincide in the Gaussian limit with uninformative priors.

In \cref{tab:results} we also show the results for the Bayes ratio, interpreted with the Jeffreys' scale as used by \cite{DES-3x2:2018}, and shown in \cref{tab:jeffreys}. 
As we can see from the table the interpretation of $R$ transitions very quickly from 
`Strong Agreement' to `Strong Tension'. To further investigate the relation between $R$ and the other metrics we plot them against each other in \cref{fig:BayesCalibration}.
This immediately highlights that the Jeffreys' scale that we use to interpret the Bayes ratio results lacks granularity 
in how it quantifies physical tensions.
Coherently across different estimators the interpretation of $R$ goes from one extreme case to the other in a probability interval that covers about one standard deviation.
\cref{fig:BayesCalibration} also clearly shows the bias of the evidence ratio toward agreement. The value of $R=1$, which separates agreement and disagreement for our choice of priors is at a probability level that roughly corresponds to $3\sigma$ (i.e.\ a probability of the discrepancy occurring by chance of $p_T \sim 0.003$).
We note that the offset between $R=1$ and $50\%$ probability events is set by the prior width and would hence change when changing the prior.
\cref{fig:BayesCalibration} also shows that the evidence ratio, interpreted with the Jeffreys' scale, would still signal a strong tension, if present, while lacking granularity in the discrimination of mildly statistically significant tensions.

\begin{figure*}
\includegraphics[width=1.7\columnwidth]{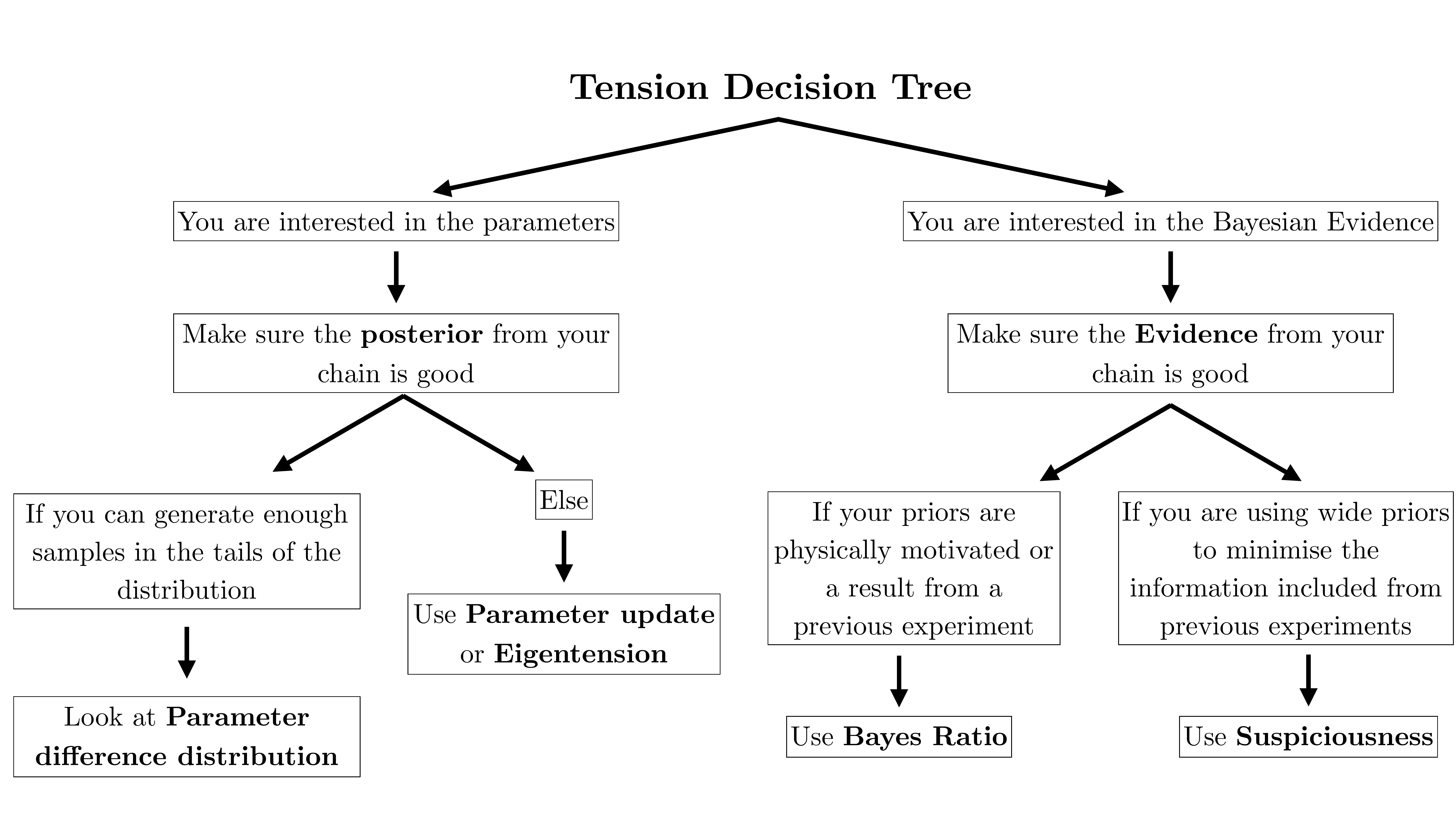}
\caption{
A practical `decision tree' to measure tension, illustrating when each tension
metric should be used. 
}
\label{fig:tree}
\end{figure*}

In \cref{sec:metrics}, we made a distinction between parameter-space methods and evidence-based methods. 
We find that all our tension metrics agree well not only amongst themselves, but also qualitatively with the a-priori Gaussian tension calculations described in \cref{sec:simulated}. This is a non-trivial result, as both the calculations and the fundamental questions that the various methods are trying to address differ.

The only exceptions to this good agreement are given by the statistically-significant $\sigma_8$ shifts where the spread between the three parameter difference estimators is smaller than the difference between them GoF loss and Suspiciousness; and the smaller a-priori shifts in $\Omega_{\rm m}$, for which the a-priori Gaussian tension estimate is smaller than the results from Eigentension and Suspiciousness. Since the input calculation used a noiseless data vector and simulated DES data vectors had noise, these disagreements are expected.
They  are likely to be caused by the noise introduced in the chains used by the tension metrics, 
and will have a more
significant impact on the small shifts.

Based on these results, we propose a methodology to quantify tension between data sets that
exploits the strengths of all the different methods, summarized by \cref{fig:tree}. Within the parameter-based approach, we recommend to generate a {\bf Monte-Carlo parameter difference distribution} and observe where the zero-difference point stands provided we have enough samples of the posterior distribution 
in its tail, as this method has no problem with non-Gaussianities, and has the advantage of providing useful visualizations in the form of confidence regions generated directly from the difference chain itself. However, if the number of samples in the tension tail is insufficient, this parameter-difference distribution will not be reliable enough to make statements about tension. In this case, either {\bf Eigentension} or {\bf parameter differences in update form} provide reliable metrics of tension. These two methods are also useful in identifying the physics behind the tension, as they provide characteristic parameter combinations along with the identified tensions lie. Since it does not offer mitigation of non-Gaussianities, we do not recommend using goodness-of-fit loss on its own, but rather as a cross-check with other metrics. 

For the Evidence-based methods, if we have a well-motivated prior, such as the posterior from a previous experiment or a physically-motivated one, we can calculate the tension using the {\bf Bayes ratio}. However, as discussed in the text, experiments such as DES and \Planck\ often choose wide priors in order to obtain posteriors that do not depend on previous experiments. The arbitrariness in the choice of width of those priors means that we cannot use the Bayes ratio, as discussed in \cref{sec:bayesr}, unless we calibrated $R$ using \cref{fig:BayesCalibration}, but that would require recalibration if any details of the analysis changed. In the case of wide and uninformative priors, the {\bf Suspiciousness} answers the same question as the Bayes ratio but correcting for the prior volume effect.
We recommend its use over the Bayes ratio in general since it has the additional desirable property of having a `tension probability' interpretation under a Gaussian approximation, without any need for calibration. 

As pointed out in \cref{fig:tree}, different methods requires reliable calculations of different quantities.
Parameter-space methods require a good estimate of the posterior, and particularly of its mean and 
covariance matrix. Evidence-based methods require a calculation of the Bayesian Evidence. Therefore, 
our choice of tension metric should inform our sampling choices, as further discussed in \cite{DES-Samplers:2020}.

\section{Application to DES Y1 and Planck} 
\label{sec:y1}

With a better understanding of the interpretation of each of the tension metrics, we now revisit the issue of consistency between the DES Y1 cosmology results and those obtained by the \Planck\ collaboration \citep{PlanckParameters:2016, PlanckParameters:2018}. 
This also serves as a worked example on real data of how tension between experiments can be fully quantified.

We choose to investigate three different combinations of DES data sets: (1) weak lensing-only constraints from \citet{Troxel:2017xyo} (2) constraints from combining the auto and cross-correlation between weak lensing and galaxy clustering, referred to as the $3\times 2$pt analysis: (3) constraints from (2) plus cross-correlation with CMB lensing, referred as the $5\times 2$pt analysis \citep{DES-5x2:2019}. 
We particularly focus in the second combination, as it provided the most powerful 
constraints from large-scale structure measured by DES alone.
For \textit{Planck} 2015 we use the small-scale ($\ell>30$) measurements of the CMB temperature power spectrum and the joint large-scale temperature and polarization data.
For \textit{Planck} 2018 we use small-scale CMB temperature, polarization and their cross-correlation measurements combined with large-scale temperature and and $E$-mode polarization data.
In doing so we follow the recommendations of the \Planck\ collaboration in the two data releases.

\begin{figure}
	\includegraphics[width=\columnwidth]{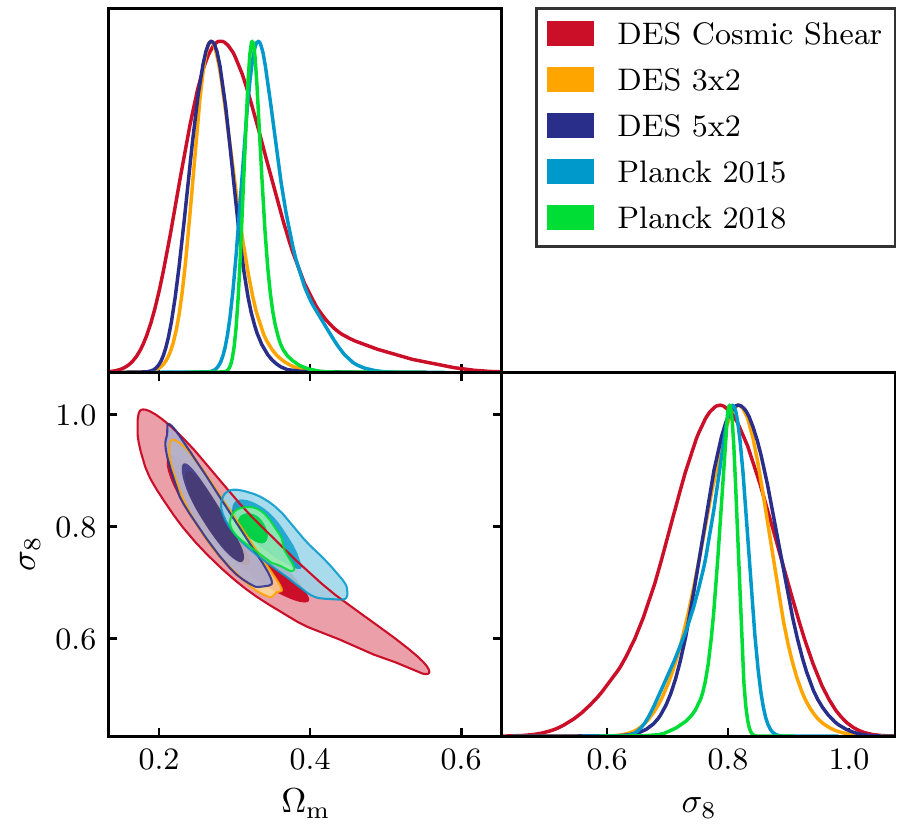}
    \caption{
    \label{fig:y1}
    $68 \%$ and $95 \%$ confidence regions of the joint marginalized posterior probability distributions for Dark Energy Survey Year 1 Cosmic Shear, $3\times 2$pt and $5\times 2$pt likelihoods, and for the \Planck\ 2015 TTTEEE likelihood. 
    }
\end{figure}

The results of parameter estimation for these data sets are shown in \cref{fig:y1} and the results of different tension estimators in \cref{tab:results_desy1}.
We highlight in the table the results that we focus our discussion on.\footnote{The reader might notice that the values of the Bayes ratio reported in~\cref{tab:results_desy1}, in particular for the case DES $3\times 2$pt vs. \Planck\ 15, differ from the values reported by \cite{DES-3x2:2018} ($R = 6.6$). This difference has been identified as originating from sampling issues in the DES Y1 analysis, as will be described in more detail in \cite{DES-Samplers:2020}.
}

\begin{figure*}
	\includegraphics[width=0.495\textwidth]{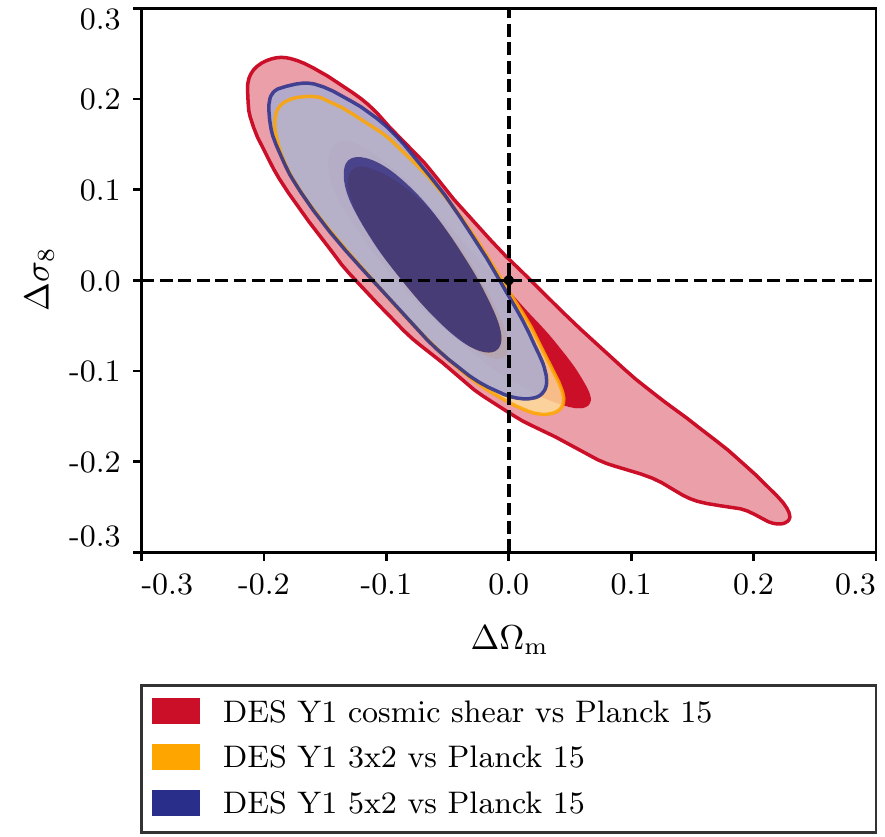}
	\includegraphics[width=0.495\textwidth]{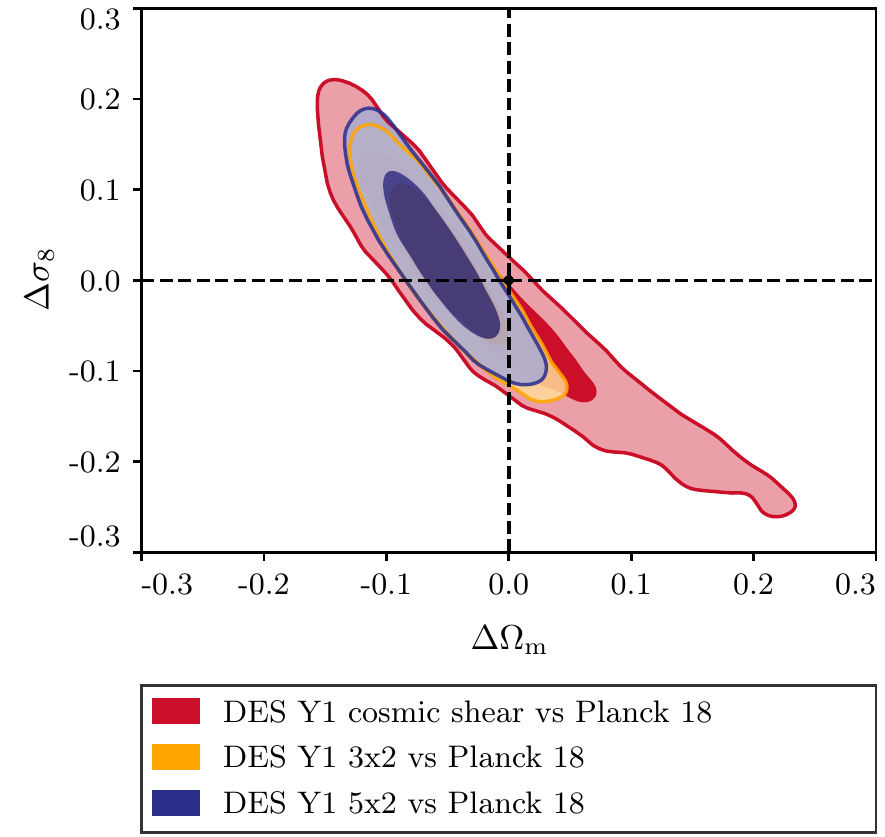}
    \caption{ \label{fig:data_shifts}
    Joint marginalized posterior distribution of the parameter differences between different DES data selections and \Planck\ 15/18. The distribution of parameter differences is used to compute the statistical significance of a parameter shift. The darker and lighter shading corresponds to the $68\%$ and $95\%$ C.L. regions respectively.
    }
\end{figure*}

We start with MCMC parameter shifts, as it is the parameter-based method that can give the most accurate value for the tension, thanks to its ability to go beyond the Gaussian approximation. In \cref{fig:data_shifts} we can see the posterior of differences between the determination of $\sigma_8$ and $\Omega_{\rm m}$ from different DES data sets and \Planck\ that clearly shows a tension that is greater than $2\sigma$. 
In \cref{tab:results_desy1} we see that in full parameter space this tension is at the $2.2\sigma$ level.
We proceed with Suspiciousness as our recommended evidence-based method which fully confirms the parameter-shift results, giving a $2.4 \pm 0.2\sigma$ tension between \Planck\ 2015 and DES $3\times 2$pt.
We note that applying both methods provides a useful cross-check of their respective results.
This moderate tension remains when \Planck\ is updated from the 2015 to the 2018 data and for DES $5\times 2$pt. 
This shows that this tension is robust to the inclusion of CMB polarization data.

\begin{table*}
\centering
\setlength{\tabcolsep}{0.6em}
\begin{tabular}{|l|cccccc}
        \hline
    \multirow{2}{*}{data set} & \multicolumn{2}{c}{Bayes ratio~~} & \multirow{2}{*}{Eigentension} & \multirow{2}{*}{GoF Loss} & MCMC/Update & \multirow{2}{*}{Suspiciousness} \\
     & $\hfill\log R\hfill$ & Interpretation & & & Param Shifts &  \\   
\hline \hline
        DES cosmic shear vs. \Planck\ 15 & $2.2 \pm 0.5$ & Substantial Agreement & $1.8\, \sigma$ & $1.3 \,\sigma$ & $1.3/1.2 \,\sigma$ & $(0.7 \pm 0.4) \,\sigma$ \\
        {\bf DES $3\times 2$pt vs. \Planck\ 15} & $1.0 \pm 0.5$ & No Evidence & $2.4\,\sigma$ & $2.7\,\sigma$ & $2.2/2.2 \,\sigma$ & $(2.4 \pm 0.2) \,\sigma$  \\
        DES $5\times 2$pt vs. \Planck\ 15 & $1.1 \pm 0.5$ & Substantial Agreement & $2.4\,\sigma$ & $2.8\,\sigma$ & $2.1/2.3 \,\sigma$ & $(2.2 \pm 0.3) \,\sigma$  \\
        DES $5\times 2$pt vs. \Planck\ 15 + lensing & $1.0 \pm 0.6$ & No Evidence & $2.4\,\sigma$ & $2.5\,\sigma$ & $2.1/2.3 \,\sigma$ &  $(2.2 \pm 0.4) \,\sigma$ \\
        DES $5\times 2$pt + \Planck\ lensing vs. \Planck\ 15 & $6.1 \pm 0.6$ & Strong Agreement & $1.6\,\sigma$ & $2.4\,\sigma$ & $1.9/2.2 \,\sigma$ & $(1.8 \pm 0.2) \,\sigma$  \\
        \hline
        DES cosmic shear vs. \Planck\ 18 & $3.3 \pm 0.4$ & Strong Agreement & $1.5\,\sigma$ & $1.0\,\sigma$ & $1.0/1.1 \,\sigma$ &  $(0.5 \pm 0.3) \,\sigma$  \\
        {\bf DES $3\times 2$pt vs. \Planck\ 18} & $2.2 \pm 0.6$ & Substantial Agreement & $2.2\,\sigma$ & $1.6\,\sigma$ & $2.0/2.3 \,\sigma$ &  $(2.4 \pm 0.2) \,\sigma$   \\
        \hline 
    \end{tabular}
\caption{
The tension between \Planck\ and different data set combinations involving DES Y1 data, calculated via the different tension metrics described in the main text. In the first column, \Planck\ refers to the combination of the TT, TE and EE likelihoods. In bold font we highlight the combinations of DES $3\times 2$pt and \Planck, as those 
are the main focus of this section. The horizontal line separates \Planck\ 2015 and 2018 data set combinations.
}
\label{tab:results_desy1}
\end{table*}

To understand the physics behind these discrepancies, it is useful to consider other methods.
Using Eigentension, we identify a single well-measured eigenmode for each DES analysis: $\sigma_8 \Omega_\mathrm{m}^{0.57}$ for the $3\times 2$pt analysis, and $\sigma_8 \Omega_\mathrm{m}^{0.58}$ in the $5\times 2$pt case. Both eigenmodes are very similar to the widely-used definition of $S_8 = \sigma_8 (\Omega_\mathrm{m}/0.3)^{0.5}$, and can be interpreted as representing the `lensing strength' arising from the large-scale structure of the late-time Universe. 
After measuring tension exclusively along this direction in parameter space, we find results that are in agreement with other methods. This shows that the moderate tension between DES and \Planck\ is found along a parameter space direction that we believe DES is robustly measuring.
Studying parameter updates of DES with respect to \Planck\ gives similar conclusions.
As discussed in the previous section and shown in \cref{fig:kl}, combining DES improves the \Planck\ determination of two parameters, the first mode projecting mostly onto $\sigma_8$ and the second onto $\Omega_{\rm m} h^2$. The first mode drives most of the tension while the shift in the second is compatible with a statistical fluctuation.
Decrease in Goodness of Fit agrees with other estimators.

The Bayes ratio interpreted on the Jeffreys' scale reports no significant tension between all data combinations that we consider.
Given the results of the previous section we can understand this as the data tension not overcoming the bias of the Bayes ratio toward agreement.
We note that the priors used for the fiducial analyses in the previous section do not coincide with the priors used in this section; we thus cannot use the previously-derived calibration of the Bayes ratio.

The mild tension we obtain between \Planck\ and DES, varying between $2 \sigma$ and $3 \sigma$, should not be overlooked. 
While this level of tension could still be a statistical fluke, it is significant enough to warrant in-depth future investigations. 
The forthcoming DES Y3 analysis, incorporating a larger fraction of the sky, 
is expected to shed light on this matter.

\section{Conclusions}
\label{sec:conclusions}
In this work, we have explored different methods to quantify consistency between two uncorrelated data sets, focusing on the comparison between DES and \Planck. The motivation
is to decide on a metric of tension between these two surveys ahead of the DES Y3 data 
release. This was done by simulating a set of DES data sets with values of cosmological parameters chosen to introduce varying levels of discrepancy with \Planck. We calculate the tension for each simulated DES data set, and compare to an a-priori Gaussian tension expected based on the known true cosmologies for the simulated data sets. While this work has been performed for the specific case of DES and \Planck, our findings about the different metrics described in
\cref{sec:results} apply to any problem of tension quantification. However, if we wanted to apply the Bayes ratio to 
a different problem with uninformative priors, the exercise of calibrating the Bayes ratio would have to be repeated.

We have found that the Bayes' ratio used in the Y1 analysis has several flaws that make 
it unsuitable for the quantitative comparison of DES and \Planck. In particular, it is proportional to the 
width of the chosen uninformative prior; it relies on the Jeffreys' scale to interpret the ratio 
of probabilities, which needs an unknown calibration that is problem-dependent (i.e. we would
need to build a table such as \cref{tab:results} in every problem to calculate the overall
calibration of the Bayes ratio); and the fact that 
we can only calculate logarithms of the probability ratio means that the Jeffreys' scale used
in the DES Y1 analysis (\cref{tab:jeffreys}) will in most cases diagnose extreme agreement or extreme tension. 

As shown in \cref{tab:results}, the other four tension metrics employed in this work --- Eigentension, GoF loss, Parameter differences, and Suspiciousness --- agree with the a-priori tension, as well as amongst themselves,
with the exceptions of small shifts in $\Omega_{\rm m}$ and large shifts in $\sigma_{8}$ discussed in \cref{sec:results}, which are likely the result of noise introduced in the simulated data vectors.
We conclude that any of the tension metrics can be used for the problem of quantifying tension between DES and \Planck, as they produce similar results. 

We use these tension metrics to re-assess the tension between DES Y1 and \Planck\ 2015, as well 
as with the latest \Planck\ 2018 results. We find, similar to our findings from the simulated analyses, that the dependence of the Evidence ratio on calibration causes the results to be inconsistent with what we see in the plots, and 
what all other tension metrics indicate. We find that there is a  $\sim 2.3 \sigma$ 
between DES and \Planck, which remains when the \Planck\ 2018 likelihood is used. It remains to be seen how this will evolve when the more powerful DES Y3 data are used. If the tension is reduced when more data are considered, we are likely looking at a statistical fluctuation. If the tension remains or increases, we could be looking at unexplained systematics in either of the surveys, or evidence of physics beyond the \LCDM\ model.

\section*{Data availability Statement}
The data underlying this article are available in the Dark Energy Survey Data Management platform, at \url{https://des.ncsa.illinois.edu}

\section*{Acknowledgements}
Funding for the DES Projects has been provided by the U.S. Department of Energy, the U.S. National Science Foundation, the Ministry of Science and Education of Spain, 
the Science and Technology Facilities Council of the United Kingdom, the Higher Education Funding Council for England, the National Center for Supercomputing 
Applications at the University of Illinois at Urbana-Champaign, the Kavli Institute of Cosmological Physics at the University of Chicago, 
the Center for Cosmology and Astro-Particle Physics at the Ohio State University,
the Mitchell Institute for Fundamental Physics and Astronomy at Texas A\&M University, Financiadora de Estudos e Projetos, 
Funda{\c c}{\~a}o Carlos Chagas Filho de Amparo {\`a} Pesquisa do Estado do Rio de Janeiro, Conselho Nacional de Desenvolvimento Cient{\'i}fico e Tecnol{\'o}gico and 
the Minist{\'e}rio da Ci{\^e}ncia, Tecnologia e Inova{\c c}{\~a}o, the Deutsche Forschungsgemeinschaft and the Collaborating Institutions in the Dark Energy Survey. 

The Collaborating Institutions are Argonne National Laboratory, the University of California at Santa Cruz, the University of Cambridge, Centro de Investigaciones Energ{\'e}ticas, 
Medioambientales y Tecnol{\'o}gicas-Madrid, the University of Chicago, University College London, the DES-Brazil Consortium, the University of Edinburgh, 
the Eidgen{\"o}ssische Technische Hochschule (ETH) Z{\"u}rich, 
Fermi National Accelerator Laboratory, the University of Illinois at Urbana-Champaign, the Institut de Ci{\`e}ncies de l'Espai (IEEC/CSIC), 
the Institut de F{\'i}sica d'Altes Energies, Lawrence Berkeley National Laboratory, the Ludwig-Maximilians Universit{\"a}t M{\"u}nchen and the associated Excellence Cluster Universe, 
the University of Michigan, NFS's NOIRLab, the University of Nottingham, The Ohio State University, the University of Pennsylvania, the University of Portsmouth, 
SLAC National Accelerator Laboratory, Stanford University, the University of Sussex, Texas A\&M University, and the OzDES Membership Consortium.

Based in part on observations at Cerro Tololo Inter-American Observatory at NSF's NOIRLab (NOIRLab Prop. ID 2012B-0001; PI: J. Frieman), which is managed by the Association of Universities for Research in Astronomy (AURA) under a cooperative agreement with the National Science Foundation.

The DES data management system is supported by the National Science Foundation under Grant Numbers AST-1138766 and AST-1536171.
The DES participants from Spanish institutions are partially supported by MICINN under grants ESP2017-89838, PGC2018-094773, PGC2018-102021, SEV-2016-0588, SEV-2016-0597, and MDM-2015-0509, some of which include ERDF funds from the European Union. IFAE is partially funded by the CERCA program of the Generalitat de Catalunya.
Research leading to these results has received funding from the European Research
Council under the European Union's Seventh Framework Program (FP7/2007-2013) including ERC grant agreements 240672, 291329, and 306478.
We  acknowledge support from the Brazilian Instituto Nacional de Ci\^encia
e Tecnologia (INCT) do e-Universo (CNPq grant 465376/2014-2).

This manuscript has been authored by Fermi Research Alliance, LLC under Contract No. DE-AC02-07CH11359 with the U.S. Department of Energy, Office of Science, Office of High Energy Physics.

PL acknowledges STFC Consolidated Grants ST/R000476/1 and ST/T000473/1. We also thank the organizers of the DES Y3 workshop: "Probing Dark Energy Observations in the Nonlinear Regime" at the
University of Michigan in Ann Arbor, where this project started.


\bibliographystyle{mnras_2author}
\bibliography{refs} 

\begin{thebibliography}{}
\makeatletter
\relax
\def\mn@urlcharsother{\let\do\@makeother \do\$\do\&\do\#\do\^\do\_\do\%\do\~}
\def\mn@doi{\begingroup\mn@urlcharsother \@ifnextchar [ {\mn@doi@}
  {\mn@doi@[]}}
\def\mn@doi@[#1]#2{\def\@tempa{#1}\ifx\@tempa\@empty \href
  {http://dx.doi.org/#2} {doi:#2}\else \href {http://dx.doi.org/#2} {#1}\fi
  \endgroup}
\def\mn@eprint#1#2{\mn@eprint@#1:#2::\@nil}
\def\mn@eprint@arXiv#1{\href {http://arxiv.org/abs/#1} {{\tt arXiv:#1}}}
\def\mn@eprint@dblp#1{\href {http://dblp.uni-trier.de/rec/bibtex/#1.xml}
  {dblp:#1}}
\def\mn@eprint@#1:#2:#3:#4\@nil{\def\@tempa {#1}\def\@tempb {#2}\def\@tempc
  {#3}\ifx \@tempc \@empty \let \@tempc \@tempb \let \@tempb \@tempa \fi \ifx
  \@tempb \@empty \def\@tempb {arXiv}\fi \@ifundefined
  {mn@eprint@\@tempb}{\@tempb:\@tempc}{\expandafter \expandafter \csname
  mn@eprint@\@tempb\endcsname \expandafter{\@tempc}}}

\bibitem[\protect\citeauthoryear{Abbott et~al.}{Abbott
  et~al.}{2016}]{Abbott:2016ktf}
Abbott T.,  et~al., 2016, \mn@doi [Mon. Not. Roy. Astron. Soc.]
  {10.1093/mnras/stw641}, 460, 1270

\bibitem[\protect\citeauthoryear{{Abbott} \& {Abdalla} et~al.,}{{Abbott}
  et~al.}{2018}]{DES-3x2:2018}
{Abbott} T.~M.~C.,  et~al. 2018, \mn@doi [\prd] {10.1103/PhysRevD.98.043526},
  \href {https://ui.adsabs.harvard.edu/abs/2018PhRvD..98d3526A} {98, 043526}

\bibitem[\protect\citeauthoryear{Abbott et~al.}{Abbott
  et~al.}{2019a}]{DES-5x2:2019}
Abbott T.,  et~al., 2019a, \mn@doi [Phys. Rev. D]
  {10.1103/PhysRevD.100.023541}, 100, 023541

\bibitem[\protect\citeauthoryear{{Abbott} \& {Alarcon} et~al.,}{{Abbott}
  et~al.}{2019b}]{DES-Combined:2018}
{Abbott} T.~M.~C.,  et~al. 2019b, \mn@doi [] {10.1103/PhysRevLett.122.171301},
  \href {https://ui.adsabs.harvard.edu/abs/2019PhRvL.122q1301A} {122, 171301}

\bibitem[\protect\citeauthoryear{{Abbott} \& {Allam} et~al.,}{{Abbott}
  et~al.}{2019c}]{DES-SNe:2019}
{Abbott} T.~M.~C.,  et~al. 2019c, \mn@doi [\apj] {10.3847/2041-8213/ab04fa},
  \href {https://ui.adsabs.harvard.edu/abs/2019ApJ...872L..30A} {872, L30}

\bibitem[\protect\citeauthoryear{Adhikari \& Huterer}{Adhikari \&
  Huterer}{2019}]{Adhikari:2018wnk}
Adhikari S.,  Huterer D.,  2019, \mn@doi [JCAP]
  {10.1088/1475-7516/2019/01/036}, 1901, 036

\bibitem[\protect\citeauthoryear{{Allen}, {Evrard}  \& {Mantz}}{{Allen}
  et~al.}{2011}]{Allen:2011}
{Allen} S.~W.,  {Evrard} A.~E.,   {Mantz} A.~B.,  2011, \mn@doi [\araa]
  {10.1146/annurev-astro-081710-102514}, \href
  {https://ui.adsabs.harvard.edu/abs/2011ARA&A..49..409A} {49, 409}

\bibitem[\protect\citeauthoryear{Amendola, Marra  \& Quartin}{Amendola
  et~al.}{2013}]{Amendola:2012wc}
Amendola L.,  Marra V.,   Quartin M.,  2013, \mn@doi [Mon. Not. Roy. Astron.
  Soc.] {10.1093/mnras/stt008}, 430, 1867

\bibitem[\protect\citeauthoryear{{Asgari} \& {Lin} et~al.,}{{Asgari}
  et~al.}{2020}]{Asgari:2020}
{Asgari} M.,  et~al. 2020, arXiv e-prints, \href
  {https://ui.adsabs.harvard.edu/abs/2020arXiv200715633A} {p. arXiv:2007.15633}

\bibitem[\protect\citeauthoryear{Box \& Cox}{Box \& Cox}{1964}]{Box:1964}
Box G. E.~P.,  Cox D.~R.,  1964, Journal of the Royal Statistical Society.
  Series B (Methodological), 26, 211

\bibitem[\protect\citeauthoryear{{Bridges} \& {Feroz} et~al.,}{{Bridges}
  et~al.}{2009}]{Bridges:2009}
{Bridges} M.,  et~al. 2009, \mn@doi [\mnras]
  {10.1111/j.1365-2966.2009.15525.x}, \href
  {https://ui.adsabs.harvard.edu/\#abs/2009MNRAS.400.1075B} {400, 1075}

\bibitem[\protect\citeauthoryear{Chac{\'o}n \& Duong}{Chac{\'o}n \&
  Duong}{2018}]{chacon2018multivariate}
Chac{\'o}n J.,  Duong T.,  2018, Multivariate Kernel Smoothing and Its
  Applications.
Chapman \& Hall/CRC Monographs on Statistics and Applied Probability, CRC Press

\bibitem[\protect\citeauthoryear{Charnock, Battye  \& Moss}{Charnock
  et~al.}{2017}]{Charnock:2017vcd}
Charnock T.,  Battye R.~A.,   Moss A.,  2017, \mn@doi [Phys. Rev.]
  {10.1103/PhysRevD.95.123535}, D95, 123535

\bibitem[\protect\citeauthoryear{Doux et~al.}{Doux et~al.}{2020}]{Doux:2020kdz}
Doux C.,  et~al., 2020, arXiv:2011.03410

\bibitem[\protect\citeauthoryear{{Efstathiou}, {Sutherland}  \&
  {Maddox}}{{Efstathiou} et~al.}{1990}]{Efstathiou:1990}
{Efstathiou} G.,  {Sutherland} W.~J.,   {Maddox} S.~J.,  1990, \mn@doi [\nat]
  {10.1038/348705a0}, \href
  {https://ui.adsabs.harvard.edu/abs/1990Natur.348..705E} {348, 705}

\bibitem[\protect\citeauthoryear{{Eisenstein}, {Seo}  \& {White}}{{Eisenstein}
  et~al.}{2007}]{Eisenstein:2007}
{Eisenstein} D.~J.,  {Seo} H.-J.,   {White} M.,  2007, \mn@doi [\apj]
  {10.1086/518755}, \href {http://adsabs.harvard.edu/abs/2007ApJ...664..660E}
  {664, 660}

\bibitem[\protect\citeauthoryear{{Elvin-Poole} \& {Crocce}
  et~al.,}{{Elvin-Poole} et~al.}{2018}]{Elvin-Poole:2018}
{Elvin-Poole} J.,  et~al. 2018, \mn@doi [\prd] {10.1103/PhysRevD.98.042006},
  \href {https://ui.adsabs.harvard.edu/abs/2018PhRvD..98d2006E} {98, 042006}

\bibitem[\protect\citeauthoryear{Feeney \& Peiris et~al.,}{Feeney
  et~al.}{2019}]{Feeney:2018mkj}
Feeney S.~M.,  et~al. 2019, \mn@doi [Phys. Rev. Lett.]
  {10.1103/PhysRevLett.122.061105}, 122, 061105

\bibitem[\protect\citeauthoryear{{Feroz}, {Hobson}  \& {Bridges}}{{Feroz}
  et~al.}{2009}]{Feroz:2008}
{Feroz} F.,  {Hobson} M.~P.,   {Bridges} M.,  2009, \mn@doi [\mnras]
  {10.1111/j.1365-2966.2009.14548.x}, \href
  {http://adsabs.harvard.edu/abs/2009MNRAS.398.1601F} {398, 1601}

\bibitem[\protect\citeauthoryear{Gelman \& Carlin et~al.,}{Gelman
  et~al.}{2004}]{Gelman:2013}
Gelman A.,  et~al. 2004, {Bayesian data analysis; 2nd ed.}.
Chapman and Hall, Boca Raton, FL, \url {https://cds.cern.ch/record/1010408}

\bibitem[\protect\citeauthoryear{{Grandis} \& {Seehars} et~al.,}{{Grandis}
  et~al.}{2016}]{Grandis:2016}
{Grandis} S.,  et~al. 2016, \mn@doi [Journal of Cosmology and Astro-Particle
  Physics] {10.1088/1475-7516/2016/05/034}, \href
  {https://ui.adsabs.harvard.edu/\#abs/2016JCAP...05..034G} {2016, 034}

\bibitem[\protect\citeauthoryear{Handley}{Handley}{2019}]{anesthetic}
Handley W.,  2019, \mn@doi [The Journal of Open Source Software]
  {10.21105/joss.01414}, 4, 1414

\bibitem[\protect\citeauthoryear{{Handley} \& {Lemos}}{{Handley} \&
  {Lemos}}{2019a}]{Handley:2019b}
{Handley} W.,  {Lemos} P.,  2019a, arXiv e-prints, \href
  {https://ui.adsabs.harvard.edu/abs/2019arXiv190306682H} {p. arXiv:1903.06682}

\bibitem[\protect\citeauthoryear{Handley \& Lemos}{Handley \&
  Lemos}{2019b}]{Handley:2019wlz}
Handley W.,  Lemos P.,  2019b, \mn@doi [Phys. Rev.]
  {10.1103/PhysRevD.100.043504}, D100, 043504

\bibitem[\protect\citeauthoryear{{Handley}, {Hobson}  \& {Lasenby}}{{Handley}
  et~al.}{2015a}]{Handley:2015a}
{Handley} W.~J.,  {Hobson} M.~P.,   {Lasenby} A.~N.,  2015a, \mn@doi [\mnras]
  {10.1093/mnrasl/slv047}, \href
  {http://adsabs.harvard.edu/abs/2015MNRAS.450L..61H} {450, L61}

\bibitem[\protect\citeauthoryear{{Handley}, {Hobson}  \& {Lasenby}}{{Handley}
  et~al.}{2015b}]{Handley:2015b}
{Handley} W.~J.,  {Hobson} M.~P.,   {Lasenby} A.~N.,  2015b, \mn@doi [\mnras]
  {10.1093/mnras/stv1911}, \href
  {http://adsabs.harvard.edu/abs/2015MNRAS.453.4384H} {453, 4384}

\bibitem[\protect\citeauthoryear{{Heymans} \& {Tr{\"o}ster} et~al.,}{{Heymans}
  et~al.}{2020}]{Heymans:2020}
{Heymans} C.,  et~al. 2020, arXiv e-prints, \href
  {https://ui.adsabs.harvard.edu/abs/2020arXiv200715632H} {p. arXiv:2007.15632}

\bibitem[\protect\citeauthoryear{{Higson} \& {Handley} et~al.,}{{Higson}
  et~al.}{2018}]{Higson2018}
{Higson} E.,  et~al. 2018, Bayesian Analysis, \href
  {https://ui.adsabs.harvard.edu/abs/2018BayAn..13..873H} {13, 873}

\bibitem[\protect\citeauthoryear{Hobson, Bridle  \& Lahav}{Hobson
  et~al.}{2002}]{Hobson:2002zf}
Hobson M.~P.,  Bridle S.~L.,   Lahav O.,  2002, \mn@doi [Mon. Not. Roy. Astron.
  Soc.] {10.1046/j.1365-8711.2002.05614.x}, 335, 377

\bibitem[\protect\citeauthoryear{{Hosoya}, {Buchert}  \& {Morita}}{{Hosoya}
  et~al.}{2004}]{Hosoya:2004}
{Hosoya} A.,  {Buchert} T.,   {Morita} M.,  2004, \mn@doi [Physical Review
  Letters] {10.1103/PhysRevLett.92.141302}, \href
  {http://adsabs.harvard.edu/abs/2004PhRvL..92n1302H} {92, 141302}

\bibitem[\protect\citeauthoryear{{Hubble}}{{Hubble}}{1929}]{Hubble:1929}
{Hubble} E.,  1929, \mn@doi [Proceedings of the National Academy of Science]
  {10.1073/pnas.15.3.168}, \href
  {https://ui.adsabs.harvard.edu/abs/1929PNAS...15..168H} {15, 168}

\bibitem[\protect\citeauthoryear{{Jeffreys}}{{Jeffreys}}{1939}]{Jeffreys:1939}
{Jeffreys} H.,  1939, {The Theory of Probability}

\bibitem[\protect\citeauthoryear{{Joachimi} \& {Taylor}}{{Joachimi} \&
  {Taylor}}{2011}]{Joachimi:2011}
{Joachimi} B.,  {Taylor} A.~N.,  2011, \mn@doi [\mnras]
  {10.1111/j.1365-2966.2011.19107.x}, \href
  {https://ui.adsabs.harvard.edu/\#abs/2011MNRAS.416.1010J} {416, 1010}

\bibitem[\protect\citeauthoryear{Joachimi et~al.}{Joachimi
  et~al.}{2020}]{Joachimi:2020abi}
Joachimi B.,  et~al., 2020, arXiv:2007.01844

\bibitem[\protect\citeauthoryear{Joudaki et~al.}{Joudaki
  et~al.}{2017}]{Joudaki:2016mvz}
Joudaki S.,  et~al., 2017, \mn@doi [Mon. Not. Roy. Astron. Soc.]
  {10.1093/mnras/stw2665}, 465, 2033

\bibitem[\protect\citeauthoryear{Joudaki \& Ferreira et~al.,}{Joudaki
  et~al.}{2020}]{Joudaki:2020shz}
Joudaki S.,  et~al. 2020

\bibitem[\protect\citeauthoryear{{Kilbinger}}{{Kilbinger}}{2015}]{Kilbinger:2015}
{Kilbinger} M.,  2015, \mn@doi [Reports on Progress in Physics]
  {10.1088/0034-4885/78/8/086901}, \href
  {https://ui.adsabs.harvard.edu/abs/2015RPPh...78h6901K} {78, 086901}

\bibitem[\protect\citeauthoryear{{Kirshner}}{{Kirshner}}{2004}]{Kirshner:2004}
{Kirshner} R.~P.,  2004, \mn@doi [Proceedings of the National Academy of
  Science] {10.1073/pnas.2536799100}, \href
  {https://ui.adsabs.harvard.edu/abs/2004PNAS..101....8K} {101, 8}

\bibitem[\protect\citeauthoryear{Krause, Eifler  et~al.}{Krause
  et~al.}{2017}]{DES-MPP:2018}
Krause E.,  Eifler T.~F.,   et~al., 2017, arXiv:1706.09359

\bibitem[\protect\citeauthoryear{Krauss \& Turner}{Krauss \&
  Turner}{1995}]{Krauss:1995yb}
Krauss L.~M.,  Turner M.~S.,  1995, \mn@doi [Gen. Rel. Grav.]
  {10.1007/BF02108229}, 27, 1137

\bibitem[\protect\citeauthoryear{Kullback \& Leibler}{Kullback \&
  Leibler}{1951}]{Kullback:1951}
Kullback S.,  Leibler R.~A.,  1951, \mn@doi [Ann. Math. Statist.]
  {10.1214/aoms/1177729694}, 22, 79

\bibitem[\protect\citeauthoryear{{Kunz}, {Trotta}  \& {Parkinson}}{{Kunz}
  et~al.}{2006}]{Kunz:2006}
{Kunz} M.,  {Trotta} R.,   {Parkinson} D.~R.,  2006, \mn@doi [\prd]
  {10.1103/PhysRevD.74.023503}, \href
  {https://ui.adsabs.harvard.edu/\#abs/2006PhRvD..74b3503K} {74, 023503}

\bibitem[\protect\citeauthoryear{Lemos \& K\"ohlinger et~al.,}{Lemos
  et~al.}{2020}]{Lemos:2019txn}
Lemos P.,  et~al. 2020, \mn@doi [Mon. Not. Roy. Astron. Soc.]
  {10.1093/mnras/staa1836}, 496, 4647

\bibitem[\protect\citeauthoryear{Lin \& Ishak}{Lin \&
  Ishak}{2017a}]{Lin:2017ikq}
Lin W.,  Ishak M.,  2017a, \mn@doi [Phys. Rev.] {10.1103/PhysRevD.96.023532},
  D96, 023532

\bibitem[\protect\citeauthoryear{Lin \& Ishak}{Lin \&
  Ishak}{2017b}]{Lin:2017bhs}
Lin W.,  Ishak M.,  2017b, \mn@doi [Phys. Rev.] {10.1103/PhysRevD.96.083532},
  D96, 083532

\bibitem[\protect\citeauthoryear{Lin \& Ishak}{Lin \&
  Ishak}{2019}]{Lin:2019zdn}
Lin W.,  Ishak M.,  2019, arXiv:1909.10991

\bibitem[\protect\citeauthoryear{Luis~Bernal \& Peacock}{Luis~Bernal \&
  Peacock}{2018}]{Bernal:2018cxc}
Luis~Bernal J.,  Peacock J.~A.,  2018, \mn@doi [JCAP]
  {10.1088/1475-7516/2018/07/002}, 1807, 002

\bibitem[\protect\citeauthoryear{{Mandelbaum}}{{Mandelbaum}}{2018}]{Mandelbaum:2018}
{Mandelbaum} R.,  2018, \mn@doi [\araa] {10.1146/annurev-astro-081817-051928},
  \href {https://ui.adsabs.harvard.edu/abs/2018ARA&A..56..393M} {56, 393}

\bibitem[\protect\citeauthoryear{Marshall, Rajguru  \& Slosar}{Marshall
  et~al.}{2006}]{Marshall:2004zd}
Marshall P.,  Rajguru N.,   Slosar A.,  2006, \mn@doi [Phys. Rev.]
  {10.1103/PhysRevD.73.067302}, D73, 067302

\bibitem[\protect\citeauthoryear{Martin \& Ringeval et~al.,}{Martin
  et~al.}{2014}]{Martin:2014lra}
Martin J.,  et~al. 2014, \mn@doi [Phys. Rev.] {10.1103/PhysRevD.90.063501},
  D90, 063501

\bibitem[\protect\citeauthoryear{Miranda, Rogozenski  \& Krause}{Miranda
  et~al.}{2020}]{Miranda:2020lpk}
Miranda V.,  Rogozenski P.,   Krause E.,  2020, arXiv:2009.14241

\bibitem[\protect\citeauthoryear{{Nicola}, {Amara}  \& {Refregier}}{{Nicola}
  et~al.}{2019}]{Nicola:2019}
{Nicola} A.,  {Amara} A.,   {Refregier} A.,  2019, \mn@doi [Journal of
  Cosmology and Astro-Particle Physics] {10.1088/1475-7516/2019/01/011}, \href
  {https://ui.adsabs.harvard.edu/\#abs/2019JCAP...01..011N} {2019, 011}

\bibitem[\protect\citeauthoryear{{Ostriker} \& {Steinhardt}}{{Ostriker} \&
  {Steinhardt}}{1995}]{Ostriker:1995}
{Ostriker} J.~P.,  {Steinhardt} P.~J.,  1995, \mn@doi [\nat]
  {10.1038/377600a0}, \href
  {https://ui.adsabs.harvard.edu/abs/1995Natur.377..600O} {377, 600}

\bibitem[\protect\citeauthoryear{{Park} \& {Rozo}}{{Park} \&
  {Rozo}}{2019}]{ParkRozo:2019}
{Park} Y.,  {Rozo} E.,  2019, arXiv e-prints, \href
  {https://ui.adsabs.harvard.edu/abs/2019arXiv190705798P} {p. arXiv:1907.05798}

\bibitem[\protect\citeauthoryear{{Peebles}}{{Peebles}}{1984}]{Peebles:1984}
{Peebles} P.~J.~E.,  1984, \mn@doi [\apj] {10.1086/162425}, \href
  {https://ui.adsabs.harvard.edu/abs/1984ApJ...284..439P} {284, 439}

\bibitem[\protect\citeauthoryear{Perlmutter et~al.}{Perlmutter
  et~al.}{1999}]{Perlmutter:1998np}
Perlmutter S.,  et~al., 1999, \mn@doi [Astrophys. J.] {10.1086/307221}, 517,
  565

\bibitem[\protect\citeauthoryear{{Planck Collaboration} \& {Ade}
  et~al.,}{{Planck Collaboration}}{2016}]{PlanckParameters:2016}
{Planck Collaboration} 2016, \mn@doi [\aap] {10.1051/0004-6361/201525830}, 594,
  A13

\bibitem[\protect\citeauthoryear{{Planck Collaboration} \& {Aghanim}
  et~al.,}{{Planck Collaboration}}{2018}]{PlanckParameters:2018}
{Planck Collaboration} 2018, preprint, \href
  {http://adsabs.harvard.edu/abs/2018arXiv180706209P} {} (\mn@eprint {arXiv}
  {1807.06209})

\bibitem[\protect\citeauthoryear{{Prat} \& {S{\'a}nchez} et~al.,}{{Prat}
  et~al.}{2018}]{Prat:2018}
{Prat} J.,  et~al. 2018, \mn@doi [\prd] {10.1103/PhysRevD.98.042005}, \href
  {https://ui.adsabs.harvard.edu/abs/2018PhRvD..98d2005P} {98, 042005}

\bibitem[\protect\citeauthoryear{Raveri \& Hu}{Raveri \&
  Hu}{2019}]{Raveri:2018wln}
Raveri M.,  Hu W.,  2019, \mn@doi [Phys. Rev.] {10.1103/PhysRevD.99.043506},
  D99, 043506

\bibitem[\protect\citeauthoryear{Raveri, Zacharegkas  \& Hu}{Raveri
  et~al.}{2020}]{Raveri:2019gdp}
Raveri M.,  Zacharegkas G.,   Hu W.,  2020, \mn@doi [Phys. Rev. D]
  {10.1103/PhysRevD.101.103527}, 101, 103527

\bibitem[\protect\citeauthoryear{Riess et~al.}{Riess
  et~al.}{1998}]{Riess:1998cb}
Riess A.~G.,  et~al., 1998, \mn@doi [Astron. J.] {10.1086/300499}, 116, 1009

\bibitem[\protect\citeauthoryear{Riess \& Casertano et~al.,}{Riess
  et~al.}{2019}]{Riess:2019cxk}
Riess A.~G.,  et~al. 2019, \mn@doi [Astrophys. J.] {10.3847/1538-4357/ab1422},
  876, 85

\bibitem[\protect\citeauthoryear{{Schuhmann}, {Joachimi}  \&
  {Peiris}}{{Schuhmann} et~al.}{2016}]{Schuhmann:2016}
{Schuhmann} R.~L.,  {Joachimi} B.,   {Peiris} H.~V.,  2016, \mn@doi [\mnras]
  {10.1093/mnras/stw738}, \href
  {https://ui.adsabs.harvard.edu/\#abs/2016MNRAS.459.1916S} {459, 1916}

\bibitem[\protect\citeauthoryear{{Seehars} \& {Amara} et~al.,}{{Seehars}
  et~al.}{2014}]{Seehars:2014}
{Seehars} S.,  et~al. 2014, \mn@doi [\prd] {10.1103/PhysRevD.90.023533}, \href
  {https://ui.adsabs.harvard.edu/\#abs/2014PhRvD..90b3533S} {90, 023533}

\bibitem[\protect\citeauthoryear{{Seehars} \& {Grandis} et~al.,}{{Seehars}
  et~al.}{2016a}]{Seerhars:2016}
{Seehars} S.,  et~al. 2016a, \mn@doi [\prd] {10.1103/PhysRevD.93.103507}, \href
  {http://adsabs.harvard.edu/abs/2016PhRvD..93j3507S} {93, 103507}

\bibitem[\protect\citeauthoryear{Seehars \& Grandis et~al.,}{Seehars
  et~al.}{2016b}]{Seehars:2015qza}
Seehars S.,  et~al. 2016b, \mn@doi [Phys. Rev.] {10.1103/PhysRevD.93.103507},
  D93, 103507

\bibitem[\protect\citeauthoryear{Skilling}{Skilling}{2006}]{Skilling:2006}
Skilling J.,  2006, \mn@doi [Bayesian Anal.] {10.1214/06-BA127}, 1, 833

\bibitem[\protect\citeauthoryear{Spiegelhalter \& Best et~al.,}{Spiegelhalter
  et~al.}{2001}]{Spiegelhalter01bayesianmeasures}
Spiegelhalter D.~J.,  et~al. 2001, Bayesian Measures of Model Complexity and
  Fit

\bibitem[\protect\citeauthoryear{Spiegelhalter \& Best et~al.,}{Spiegelhalter
  et~al.}{2002}]{Spiegelhalter:2002}
Spiegelhalter D.~J.,  et~al. 2002, Journal of the Royal Statistical Society
  Series B, 64, 583

\bibitem[\protect\citeauthoryear{{The Dark Energy Survey Collaboration}}{{The
  Dark Energy Survey Collaboration}}{2005}]{DES:2005}
{The Dark Energy Survey Collaboration} 2005, arXiv e-prints, \href
  {https://ui.adsabs.harvard.edu/abs/2005astro.ph.10346T} {pp
  astro--ph/0510346}

\bibitem[\protect\citeauthoryear{{The Dark Energy Survey Collaboration} \&
  {Abbott} et~al.,}{{The Dark Energy Survey
  Collaboration}}{2017}]{DES-BAO:2017}
{The Dark Energy Survey Collaboration} 2017, arXiv e-prints, \href
  {https://ui.adsabs.harvard.edu/abs/2017arXiv171206209T} {p. arXiv:1712.06209}

\bibitem[\protect\citeauthoryear{{The Dark Energy Survey Collaboration}}{{The
  Dark Energy Survey Collaboration}}{2020}]{DES-Samplers:2020}
{The Dark Energy Survey Collaboration} 2020, In preparation

\bibitem[\protect\citeauthoryear{{To} \& {Krause} et~al.,}{{To}
  et~al.}{2020}]{2020arXiv201001138T}
{To} C.,  et~al. 2020, arXiv e-prints, \href
  {https://ui.adsabs.harvard.edu/abs/2020arXiv201001138T} {p. arXiv:2010.01138}

\bibitem[\protect\citeauthoryear{Troxel et~al.}{Troxel
  et~al.}{2018}]{Troxel:2017xyo}
Troxel M.,  et~al., 2018, \mn@doi [Phys. Rev. D] {10.1103/PhysRevD.98.043528},
  98, 043528

\bibitem[\protect\citeauthoryear{{Verde}, {Protopapas}  \& {Jimenez}}{{Verde}
  et~al.}{2013}]{Verde:2013}
{Verde} L.,  {Protopapas} P.,   {Jimenez} R.,  2013, \mn@doi [Physics of the
  Dark Universe] {10.1016/j.dark.2013.09.002}, \href
  {http://adsabs.harvard.edu/abs/2013PDU.....2..166V} {2, 166}

\bibitem[\protect\citeauthoryear{Wu \& Motloch et~al.,}{Wu
  et~al.}{2020}]{Wu:2020nxz}
Wu W.~K.,  et~al. 2020, \mn@doi [Phys. Rev. D] {10.1103/PhysRevD.102.023510},
  102, 023510

\bibitem[\protect\citeauthoryear{Zuntz \& Paterno et~al.,}{Zuntz
  et~al.}{2015}]{Zuntz:2014csq}
Zuntz J.,  et~al. 2015, \mn@doi [Astron. Comput.]
  {10.1016/j.ascom.2015.05.005}, 12, 45

\makeatother
\end{thebibliography}


\appendix
\section {Dark Energy Survey data}
\label{sec:appendix}

The Dark Energy Survey \citep[DES,][]{DES:2005,Abbott:2016ktf} is a six-year survey that has observed over 5000 ${\rm deg}^2$ in five filters ($grizY$) and has probed redshifts up to $z \sim 1.3$. It has also used time-domain 
to measure several thousand type Ia supernovae (SNe Ia). DES can constrain cosmological parameters in several ways: It can use these SNe Ia, and treat them as standarizable candles to constrain cosmology through their redshift-luminosity relation, usually referred to as Hubble Diagram \citep{Hubble:1929, Kirshner:2004}; it can use the distribution of galaxies to measure the Baryon Acoustic Oscillation (BAO) feature which was imprinted by sound waves at the recombination era ($z \sim 1100$), and which serves as a standard ruler \citep{Eisenstein:2007}; it can use the abundance of galaxy clusters, the largest gravitationally-bound structures in the Universe \citep{Allen:2011}; it can use the distribution of galaxies to measure the dark matter density distribution, under the assumption of some bias relating the two, called galaxy clustering; and it can measure the distortion of light by intervening matter along the line of sight, referred to as gravitational lensing \citep{Mandelbaum:2018}. When the matter distribution distorting the path of light is the large-scale structure of the Universe, the effect is called cosmic shear \citep{Kilbinger:2015}. Because in this case distortions are too small to be detected for individual galaxies, they are detected through correlations in the shapes and position of galaxies images.

Using data from the first year of observations (Y1), the DES collaboration has already reported constraints on cosmology from BAO \citep{DES-BAO:2017}, galaxy clustering \citep{Elvin-Poole:2018}, cosmic shear \citep{Troxel:2017xyo}, the cross-correlation of galaxy clustering and cosmic shear, referred to as galaxy--galaxy lensing \citep*{Prat:2018}, and as a main result, the combination of the two-point functions from cosmic shear, galaxy clustering, and galaxy--galaxy lensing, henceforth referred
to as `$3\times 2$pt' \citep{DES-3x2:2018}. In addition, using data from three years of observations (Y3), DES has also constrained cosmology from SNe Ia \citep{DES-SNe:2019},
and galaxy clusters \citep{2020arXiv201001138T}. 
However, as described in \cite{DES-Combined:2018}, the most powerful constraints from future DES data releases will come from combinations of the different probes, as these can break degeneracies in parameter constraints and significantly increase accuracy. 

\begin{table}
\begin{center}
\begin{tabular}{| c  c |}
\hline
Parameter & Prior \\  
\hline 
\multicolumn{2}{|c|}{{\bf Cosmology}} \\
$\Omega_{\rm m}$  &  flat (0.1, 0.9)  \\ 
$A_{\rm s}$ &  flat ($5\times 10^{-10},5\times 10^{-9}$)  \\ 
$n_{\rm s}$ &  flat (0.87, 1.07)  \\
$\Omega_{\rm b}$ &  flat (0.03, 0.07)  \\
$h$  &  flat (0.55, 0.90)   \\
$\Omega_\nu h^2$  & flat($5\times 10^{-4}$,$10^{-2}$) \\
\hline
\multicolumn{2}{|c|}{{\bf Lens Galaxy Bias}} \\
$b_{i} (i=1,5)$   & flat (0.8, 3.0) \\
\hline
\multicolumn{2}{|c|}{{\bf Intrinsic Alignment}} \\
$A_{\rm IA}$   & flat ($-5,5$) \\
$\eta_{\rm IA}$   & flat ($-5,5$) \\
\hline
\multicolumn{2}{|c|}{{\bf Lens photo-$z$ shift (red sequence)}} \\
$\Delta z^1_{\rm l}$  & Gauss ($0.0, 0.007$) \\
$\Delta z^2_{\rm l}$  & Gauss ($0.0, 0.007$) \\
$\Delta z^3_{\rm l}$  & Gauss ($0.0, 0.006$) \\
$\Delta z^4_{\rm l}$  & Gauss ($0.0, 0.01$) \\
$\Delta z^5_{\rm l}$  & Gauss ($0.0, 0.01$) \\
\hline
\multicolumn{2}{|c|}{{\bf Source photo-$z$ shift}} \\
$\Delta z^1_{\rm s}$  & Gauss ($0.0, 0.016$) \\
$\Delta z^2_{\rm s}$  & Gauss ($0.0, 0.013$) \\
$\Delta z^3_{\rm s}$  & Gauss ($0.0, 0.011$) \\
$\Delta z^4_{\rm s}$  & Gauss ($0.0, 0.022$) \\
\hline
\multicolumn{2}{|c|}{{\bf Shear calibration}} \\
$m^{i} (i=1,4)$ & Gauss ($0.0, 0.023$)\\
\hline
\end{tabular}
\caption{\label{tab:params}
Cosmological and nuisance parameters and their priors used in this analysis. 
}
\end{center}
\end{table}

We adopt the same priors used in the DES Y1 analysis, shown in \cref{tab:params}.

\section*{Affiliations}
$^{1}$ Department of Physics and Astronomy, Pevensey Building, University of Sussex, Brighton, BN1 9QH, UK\\
$^{2}$ Department of Physics \& Astronomy, University College London, Gower Street, London, WC1E 6BT, UK\\
$^{3}$ Kavli Institute for Cosmological Physics, University of Chicago, Chicago, IL 60637, USA\\
$^{4}$ Department of Physics, Carnegie Mellon University, Pittsburgh, Pennsylvania 15312, USA\\
$^{5}$ Kavli Institute for the Physics and Mathematics of the Universe (WPI), UTIAS, The University of Tokyo, Kashiwa, Chiba 277-8583, Japan\\
$^{6}$ Department of Astronomy and Astrophysics, University of Chicago, Chicago, IL 60637, USA\\
$^{7}$ Department of Physics, University of Michigan, Ann Arbor, MI 48109, USA\\
$^{8}$ Institute for Astronomy, University of Edinburgh, Edinburgh EH9 3HJ, UK\\
$^{9}$ Instituto de Astrof\'{\i}sica e Ci\^{e}ncias do Espa\c{c}o, Faculdade de Ci\^{e}ncias, Universidade de Lisboa, 1769-016 Lisboa, Portugal\\
$^{10}$ Perimeter Institute for Theoretical Physics, 31 Caroline St. North, Waterloo, ON N2L 2Y5, Canada\\
$^{11}$ Center for Cosmology and Astro-Particle Physics, The Ohio State University, Columbus, OH 43210, USA\\
$^{12}$ Institute of Physics, Laboratory of Astrophysics, \'Ecole Polytechnique F\'ed\'erale de Lausanne (EPFL), Observatoire de Sauverny, 1290 Versoix, Switzerland\\
$^{13}$ Physics Department, 2320 Chamberlin Hall, University of Wisconsin-Madison, 1150 University Avenue Madison, WI  53706-1390\\
$^{14}$ Lawrence Berkeley National Laboratory, 1 Cyclotron Road, Berkeley, CA 94720, USA\\
$^{15}$ Department of Physics and Astronomy, University of Pennsylvania, Philadelphia, PA 19104, USA\\
$^{16}$ Department of Physics, Stanford University, 382 Via Pueblo Mall, Stanford, CA 94305, USA\\
$^{17}$ Kavli Institute for Particle Astrophysics \& Cosmology, P. O. Box 2450, Stanford University, Stanford, CA 94305, USA\\
$^{18}$ SLAC National Accelerator Laboratory, Menlo Park, CA 94025, USA\\
$^{19}$ Department of Physics, University of Oxford, Denys Wilkinson Building, Keble Road, Oxford OX1 3RH, UK\\
$^{20}$ Jodrell Bank Center for Astrophysics, School of Physics and Astronomy, University of Manchester, Oxford Road, Manchester, M13 9PL, UK\\
$^{21}$ Department of Astronomy/Steward Observatory, University of Arizona, 933 North Cherry Avenue, Tucson, AZ 85721-0065, USA\\
$^{22}$ Department of Applied Mathematics and Theoretical Physics, University of Cambridge, Cambridge CB3 0WA, UK\\
$^{23}$ Departamento de F\'isica Matem\'atica, Instituto de F\'isica, Universidade de S\~ao Paulo, CP 66318, S\~ao Paulo, SP, 05314-970, Brazil\\
$^{24}$ Laborat\'orio Interinstitucional de e-Astronomia - LIneA, Rua Gal. Jos\'e Cristino 77, Rio de Janeiro, RJ - 20921-400, Brazil\\
$^{25}$ Fermi National Accelerator Laboratory, P. O. Box 500, Batavia, IL 60510, USA\\
$^{26}$ Instituto de Fisica Teorica UAM/CSIC, Universidad Autonoma de Madrid, 28049 Madrid, Spain\\
$^{27}$ Institute of Cosmology and Gravitation, University of Portsmouth, Portsmouth, PO1 3FX, UK\\
$^{28}$ CNRS, UMR 7095, Institut d'Astrophysique de Paris, F-75014, Paris, France\\
$^{29}$ Sorbonne Universit\'es, UPMC Univ Paris 06, UMR 7095, Institut d'Astrophysique de Paris, F-75014, Paris, France\\
$^{30}$ Instituto de Astrofisica de Canarias, E-38205 La Laguna, Tenerife, Spain\\
$^{31}$ Universidad de La Laguna, Dpto. AstrofÃ­sica, E-38206 La Laguna, Tenerife, Spain\\
$^{32}$ Center for Astrophysical Surveys, National Center for Supercomputing Applications, 1205 West Clark St., Urbana, IL 61801, USA\\
$^{33}$ Department of Astronomy, University of Illinois at Urbana-Champaign, 1002 W. Green Street, Urbana, IL 61801, USA\\
$^{34}$ Institut de F\'{\i}sica d'Altes Energies (IFAE), The Barcelona Institute of Science and Technology, Campus UAB, 08193 Bellaterra (Barcelona) Spain\\
$^{35}$ Institut d'Estudis Espacials de Catalunya (IEEC), 08034 Barcelona, Spain\\
$^{36}$ Institute of Space Sciences (ICE, CSIC),  Campus UAB, Carrer de Can Magrans, s/n,  08193 Barcelona, Spain\\
$^{37}$ University of Nottingham, School of Physics and Astronomy, Nottingham NG7 2RD, UK\\
$^{38}$ Astronomy Unit, Department of Physics, University of Trieste, via Tiepolo 11, I-34131 Trieste, Italy\\
$^{39}$ INAF-Osservatorio Astronomico di Trieste, via G. B. Tiepolo 11, I-34143 Trieste, Italy\\
$^{40}$ Institute for Fundamental Physics of the Universe, Via Beirut 2, 34014 Trieste, Italy\\
$^{41}$ School of Mathematics and Physics, University of Queensland,  Brisbane, QLD 4072, Australia\\
$^{42}$ Centro de Investigaciones Energ\'eticas, Medioambientales y Tecnol\'ogicas (CIEMAT), Madrid, Spain\\
$^{43}$ Department of Physics, IIT Hyderabad, Kandi, Telangana 502285, India\\
$^{44}$ Jet Propulsion Laboratory, California Institute of Technology, 4800 Oak Grove Dr., Pasadena, CA 91109, USA\\
$^{45}$ Department of Physics, The Ohio State University, Columbus, OH 43210, USA\\
$^{46}$ Santa Cruz Institute for Particle Physics, Santa Cruz, CA 95064, USA\\
$^{47}$ Department of Astronomy, University of Michigan, Ann Arbor, MI 48109, USA\\
$^{48}$ Institute of Theoretical Astrophysics, University of Oslo. P.O. Box 1029 Blindern, NO-0315 Oslo, Norway\\
$^{49}$ Institute of Astronomy, University of Cambridge, Madingley Road, Cambridge CB3 0HA, UK\\
$^{50}$ Kavli Institute for Cosmology, University of Cambridge, Madingley Road, Cambridge CB3 0HA, UK\\
$^{51}$ Observat\'orio Nacional, Rua Gal. Jos\'e Cristino 77, Rio de Janeiro, RJ - 20921-400, Brazil\\
$^{52}$ Department of Astronomy, University of Geneva, ch. d'\'Ecogia 16, CH-1290 Versoix, Switzerland\\
$^{53}$ Faculty of Physics, Ludwig-Maximilians-Universit\"at, Scheinerstr. 1, 81679 Munich, Germany\\
$^{54}$ Max Planck Institute for Extraterrestrial Physics, Giessenbachstrasse, 85748 Garching, Germany\\
$^{55}$ Center for Astrophysics $\vert$ Harvard \& Smithsonian, 60 Garden Street, Cambridge, MA 02138, USA\\
$^{56}$ George P. and Cynthia Woods Mitchell Institute for Fundamental Physics and Astronomy, and Department of Physics and Astronomy, Texas A\&M University, College Station, TX 77843,  USA\\
$^{57}$ Department of Astronomy, The Ohio State University, Columbus, OH 43210, USA\\
$^{58}$ Radcliffe Institute for Advanced Study, Harvard University, Cambridge, MA 02138\\
$^{59}$ Department of Astrophysical Sciences, Princeton University, Peyton Hall, Princeton, NJ 08544, USA\\
$^{60}$ Instituci\'o Catalana de Recerca i Estudis Avan\c{c}ats, E-08010 Barcelona, Spain\\
$^{61}$ School of Physics and Astronomy, University of Southampton,  Southampton, SO17 1BJ, UK\\
$^{62}$ Computer Science and Mathematics Division, Oak Ridge National Laboratory, Oak Ridge, TN 37831\\
$^{63}$ Department of Physics, Duke University Durham, NC 27708, USA\\
$^{64}$ Universit\"ats-Sternwarte, Fakult\"at f\"ur Physik, Ludwig-Maximilians Universit\"at M\"unchen, Scheinerstr. 1, 81679 M\"unchen, Germany\\


\bsp	
\label{lastpage}
\end{document}